\documentclass[
11pt, % The default document font size, options: 10pt, 11pt, 12pt
english, % ngerman for German
singlespacing, % Single line spacing, alternatives: onehalfspacing or doublespacing
headsepline, % Uncomment to get a line under the header
]{MastersDoctoralThesis} % The class file specifying the document structure
\usepackage{multirow}
\usepackage[utf8]{inputenc} % Required for inputting international characters
\usepackage[T1]{fontenc} % Output font encoding for international characters

\usepackage{subfig}
\usepackage{mathpazo} % Use the Palatino font by default

\usepackage[backend=bibtex,style=authoryear,natbib=true]{biblatex} % Use the bibtex backend with the authoryear citation style (which resembles APA)

\usepackage[autostyle=true]{csquotes} % Required to generate language-dependent quotes in the bibliography

\thesistitle{Detection of Doctored Speech: Towards an End-to-End Parametric Learn-able Filter Approach} % Your thesis title, this is used in the title and abstract, print it elsewhere with \ttitle
\supervisor{Dr. Saket \textsc{Anand}} % Your supervisor's name, this is used in the title page, print it elsewhere with \supname
\supervisora{Dr. Aanchan \textsc{Mohan}}
\examiner{} % Your examiner's name, this is not currently used anywhere in the template, print it elsewhere with \examname
\degree{Masters of Technology} % Your degree name, this is used in the title page and abstract, print it elsewhere with \degreename
\author{Rohit \textsc{Arora}} % Your name, this is used in the title page and abstract, print it elsewhere with \authorname
\addresses{} % Your address, this is not currently used anywhere in the template, print it elsewhere with \addressname

\subject{Biological Sciences} % Your subject area, this is not currently used anywhere in the template, print it elsewhere with \subjectname
\keywords{} % Keywords for your thesis, this is not currently used anywhere in the template, print it elsewhere with \keywordnames
\university{\href{https://www.iiitd.ac.in/}{Indraprastha Institute of Information Technology,New Delhi}} % Your university's name and URL, this is used in the title page and abstract, print it elsewhere with \univname
\department{\href{https://cse.iiitd.ac.in/}{Department of Computer Science}} % Your department's name and URL, this is used in the title page and abstract, print it elsewhere with \deptname
\group{\href{https://cse.iiitd.ac.in/}{General Category}} % Your research group's name and URL, this is used in the title page, print it elsewhere with \groupname
\faculty{\href{https://www.iiitd.edu.in/~anands}{Saket Anand}} % Your faculty's name and URL, this is used in the title page and abstract, print it elsewhere with \facname

\AtBeginDocument{
\hypersetup{pdftitle=\ttitle} % Set the PDF's title to your title
\hypersetup{pdfauthor=\authorname} % Set the PDF's author to your name
\hypersetup{pdfkeywords=\keywordnames} % Set the PDF's keywords to your keywords
}

\begin{document}

\frontmatter % Use roman page numbering style (i, ii, iii, iv...) for the pre-content pages

\pagestyle{plain} % Default to the plain heading style until the thesis style is called for the body content

%----------------------------------------------------------------------------------------
%	TITLE PAGE
%----------------------------------------------------------------------------------------

\begin{titlepage}
\begin{center}

\vspace*{.06\textheight}
{\scshape\LARGE \univname\par}\vspace{1.5cm} % University name
\textsc{\Large Masters Thesis}\\[0.5cm] % Thesis type

\HRule \\[0.4cm] % Horizontal line
{\huge \bfseries \ttitle\par}\vspace{0.4cm} % Thesis title
\HRule \\[1.5cm] % Horizontal line
 
\begin{minipage}[t]{0.4\textwidth}
\begin{flushleft} \large
\emph{Author:}\\
\href{https://www.linkedin.com/in/rohit18115/}{\authorname} % Author name - remove the \href bracket to remove the link
\end{flushleft}
\end{minipage}
\begin{minipage}[t]{0.4\textwidth}
\begin{flushright} \large
\emph{Guide:} \\
\href{https://www.iiitd.edu.in/~anands}{\supname}\\ % Supervisor name - remove the \href bracket to remove the link
\emph{Co - Guide:} \\
\href{https://www.linkedin.com/in/aanchan-mohan-0769995/}{\supnamea} % Supervisor name - remove the \href bracket to remove the link  
\end{flushright}
\end{minipage}\\[3cm]
 
\vfill

\large \textit{A thesis submitted in fulfillment of the requirements\\ for the degree of \degreename}\\[0.3cm] % University requirement text
\textit{in the}\\[0.4cm]
\groupname\\\deptname\\[2cm] % Research group name and department name
 
\vfill

{\large \today}\\[4cm] % Date

\vfill
\end{center}
\end{titlepage}

%----------------------------------------------------------------------------------------
%	DECLARATION PAGE
%----------------------------------------------------------------------------------------

\begin{declaration}
\addchaptertocentry{\authorshipname} % Add the declaration to the table of contents
\noindent I, \authorname, declare that this thesis titled, \enquote{\ttitle} and the work presented in it are my own. I confirm that:

\begin{itemize} 
\item This work was done wholly or mainly while in candidature for a research degree at this University.
\item Where any part of this thesis has previously been submitted for a degree or any other qualification at this University or any other institution, this has been clearly stated.
\item Where I have consulted the published work of others, this is always clearly attributed.
\item Where I have quoted from the work of others, the source is always given. With the exception of such quotations, this thesis is entirely my own work.
\item I have acknowledged all main sources of help.
\item Where the thesis is based on work done by myself jointly with others, I have made clear exactly what was done by others and what I have contributed myself.\\
\end{itemize}
 
\noindent Signed:\\
\rule[0.5em]{25em}{0.5pt} % This prints a line for the signature
 
\noindent Date:\\
\rule[0.5em]{25em}{0.5pt} % This prints a line to write the date
\end{declaration}

\cleardoublepage

%----------------------------------------------------------------------------------------
%	QUOTATION PAGE
%----------------------------------------------------------------------------------------

\vspace*{0.2\textheight}

\noindent\enquote{\itshape When building habits, choose consistency over content. The best book is the one you can’t put down. The best exercise is the one you enjoy doing every day. The best health food is the one you find tasty. The best work is the work you’d do for free.}\bigbreak

\hfill Naval Ravikant

%----------------------------------------------------------------------------------------
%	ABSTRACT PAGE
%----------------------------------------------------------------------------------------

\begin{abstract}
\addchaptertocentry{\abstractname} % Add the abstract to the table of contents

{\hskip 2em}The Automatic Speaker Verification (ASV) systems have potential in biometrics applications for logical control access and authentication. A lot of things happen to be at stake if the ASV system is compromised. The questions that this dissertation explores are what are the weaknesses in the spoofing system? (or what parts of the natural speech, the spoofing attacks cannot replicate?). What are the spectral features that are best suited for distinguishing spoofed and natural speech? How efficient are the traditional handcrafted features as compared to the End-to-End(E2E) deep learning-based architectures in detecting such attacks? What are the relevant frequency regions in the spoofed samples that help distinguish between spoofed and natural speech? 

% {\hskip 2em} Through the process of exploring these questions we ended up building a wavelet based E2E parametric learnable Filter approach for spoof detection.

{\hskip 2em} To explore these questions, the preliminary work of this dissertation presents a comparative analysis on the wavelet and MFCC based state-of-the-art spoof detection techniques developed in these papers, respectively (\cite{7472724}) (\cite{alam2016}). Two datasets are used for evaluation. The results on ASVspoof 2015 justify our inclination towards wavelet-based features instead of MFCC features.  The experiments on the ASVspoof 2019 database show the lack of credibility of the traditional handcrafted features and give us more reason to progress towards using end-to-end deep neural networks and more recent techniques.

{\hskip 2em} For the subsequent few experiments, we use Sincnet architecture as our baseline. We get E2E deep learning models, which we call WSTnet and CWTnet, respectively, by replacing the Sinc layer with the Wavelet Scattering and Continuous wavelet transform layers. The results obtained from the score level fusion of our models: CWTnet and WSnet, with that of Sincnet, are encouraging and point to the fact that spectral diversity at the input feature level is an asset. The fusion model achieved 62\% and 17\% relative improvement over traditional handcrafted models and our Sincnet baseline when evaluated on the modern spoofing attacks in ASVspoof 2019. 

{\hskip 2em} The final scale distribution and the number of scales used in CWTnet are far from optimal for the task at hand. Our main motto here was to fine-tune the scale parameter to get an insight into the frequency regions responsible for distinguishing spoofed and natural speech. But manual fine-tuning and cross-validation would be very computationally and time expensive. So to solve this problem, we replaced the CWT layer with a Wavelet Deconvolution(WD) (\cite{NIPS2018_7711}) layer in our CWTnet architecture. This layer calculates the Discrete-Continuous Wavelet Transform similar to the CWTnet but also optimizes the scale parameter using back-propagation. WD layer calculates the transform on the input signal during the forward pass, gives the resultant features to the further layers of the network. It computes the gradients of the loss function with respect to the scale parameters during back-propagation. 

The WDnet model achieved $26\%$ and $7\%$ relative improvement over CWTnet and Sincnet model respectively when evaluated over ASVspoof 2019 dataset. This shows that more generalized features are extracted as compared to the features extracted by CWTnet as only the most important and relevant frequency regions are focused upon.

\end{abstract}

%----------------------------------------------------------------------------------------
%	ACKNOWLEDGEMENTS
%----------------------------------------------------------------------------------------

\begin{acknowledgements}
\addchaptertocentry{\acknowledgementname} % Add the acknowledgements to the table of contents
There are many people I would like to express gratitude for their contribution, both directly and indirectly, to this thesis. Starting chronologically, I am thankful for the era I was born in. It seems magical, fascinating, and logical all at the same time. No phonetic combinations of vowels and consonants are enough to show my appreciation for my family, my dad, my mom, and my brother, Anuj Arora, Anita Arora, and Anmol Arora. They instilled a relentless work ethic, perseverance, the ability to love unconditionally and showed me the power of being a delusional optimist. I owe my life to them.

Studying at IIITD was a tipping point for me. I would like to thank IIITD for fostering such a focused research environment. Watching the faculty members and students pour their heart and soul into solving real-world problems fueled me to take courses that interested me instead of easy ones and would fetch me the most marks. Machine Learning(ML), Deep Learning(DL), and Speech recognition and Understanding(SRU) were the toughest yet the most intriguing courses that I opted for during my master's tenure. All three of these courses were taught by my supervisor Dr. Saket Anand. After attending the DL and SRU course for a few weeks only, I was sure of doing my thesis on a topic at the intersection of DL and speech under him. I am grateful to him for introducing me to such an exciting field of Speech Spoof Detection. In addition to guiding me in the thesis, he gave me enough freedom to guide my project, letting me try my strengths, and through this, I learned to take ownership of my work and the skill of self-accountability. He also provided me with computing resources to carry out my experiments and gave me space in his laboratory. Having my own workspace was a big thing for me, and I appreciate him for that.

To my co-supervisor Dr. Aanchan Mohan. I thought I was a hard worker until I met him.  Being a lead data scientist at Global Relay, a father to a newborn, and with many other unspoken responsibilities, he somehow managed to spend hours with me debugging my code and teaching me good research practices. I will forever be grateful to him for giving his sincere efforts in helping me not only with my thesis but with my attempts at publishing conference papers, for giving me opportunities to fail at times, and for being patient. Through him, I got the rare opportunity to see what happens under the hood and how an expert like him approaches a problem. I know these words are not enough to thank him for his guidance, and I would like to work with him in the future too.

Finally, to my friends. The process of bringing this thesis to fruition was a rigorous and oftentimes difficult journey. The support and encouragement I received daily from my closest friends are what encouraged and enlightened me daily to keep at it. To Rachita Nayyar for being my sounding board and making ML assignments more enjoyable. To Himanshi Singh for her constant support and for helping me be more truthful to myself. To Zeenat Mirza, for making work from home more tolerable and proving that distance is only a number when making unbreakable bonds. To Sudatta Bhattacharya, for showing that curiosity can make anything more exciting and fun. I would also like to thank Gee, Anuparna, Saswati, Vishal, Akanksha, etc. (the list goes on) for being so supportive and part of my journey. 

Finally, I would like to convey my gratitude to my thesis committee members for allowing me the opportunity to present my work and for their suggestions.

\end{acknowledgements}

%----------------------------------------------------------------------------------------
%	LIST OF CONTENTS/FIGURES/TABLES PAGES
%----------------------------------------------------------------------------------------

\tableofcontents % Prints the main table of contents

\listoffigures % Prints the list of figures

\listoftables % Prints the list of tables

%----------------------------------------------------------------------------------------
%	ABBREVIATIONS
%----------------------------------------------------------------------------------------

\begin{abbreviations}{ll} % Include a list of abbreviations (a table of two columns)

\textbf{ASVspoof} & \textbf{A}utomatic \textbf{S}peaker \textbf{V}erification Spoofing and Countermeasure Challenge \\
\textbf{VC} & \textbf{V}oice \textbf{C}onversion \\
\textbf{SS} & \textbf{S}peech \textbf{S}ynthesis \\
\textbf{SSD} & \textbf{S}poofed \textbf{S}peech \textbf{D}etection \\
\textbf{EER} & \textbf{E}qual \textbf{E}rror \textbf{R}ate \\
\textbf{t-DCF} & \textbf{t}andem \textbf{D}etection \textbf{C}ost \textbf{F}fucntion \\
\textbf{DTW} & \textbf{D}ynamic \textbf{T}ime \textbf{W}arping \\
\textbf{WD} & \textbf{W}avelet  \textbf{D}econvolution \\
\textbf{CWT} & \textbf{C}ontinuous \textbf{W}avelet  \textbf{T}ansform \\
\textbf{WST} & \textbf{W}avelet \textbf{S}attering \textbf{T}ransform\\
\textbf{MFCC} & \textbf{M}el \textbf{F}requency \textbf{C}epstral \textbf{C}oefficient \\
\textbf{MWPC} & \textbf{M}el \textbf{W}avelet \textbf{P}packet \textbf{C}oefficients \\

\end{abbreviations}

%----------------------------------------------------------------------------------------
%	PHYSICAL CONSTANTS/OTHER DEFINITIONS
%----------------------------------------------------------------------------------------

% \begin{constants}{lr@{${}={}$}l} % The list of physical constants is a three column table

% % The \SI{}{} command is provided by the siunitx package, see its documentation for instructions on how to use it

% Speed of Light & $c_{0}$ & \SI{2.99792458e8}{\meter\per\second} (exact)\\
% %Constant Name & $Symbol$ & $Constant Value$ with units\\

% \end{constants}

%----------------------------------------------------------------------------------------
%	SYMBOLS
%----------------------------------------------------------------------------------------

% \begin{symbols}{lll} % Include a list of Symbols (a three column table)

% $a$ & distance & \si{\meter} \\
% $P$ & power & \si{\watt} (\si{\joule\per\second}) \\
% %Symbol & Name & Unit \\

% \addlinespace % Gap to separate the Roman symbols from the Greek

% $\omega$ & angular frequency & \si{\radian} \\

% \end{symbols}

%----------------------------------------------------------------------------------------
%	DEDICATION
%----------------------------------------------------------------------------------------

\dedicatory{I dedicate this to my parents and brother for supporting and inspiring me.} 

%----------------------------------------------------------------------------------------
%	THESIS CONTENT - CHAPTERS
%----------------------------------------------------------------------------------------

\mainmatter % Begin numeric (1,2,3...) page numbering

\pagestyle{thesis} % Return the page headers back to the "thesis" style

% Include the chapters of the thesis as separate files from the Chapters folder
% Uncomment the lines as you write the chapters

% Chapter 1

\chapter{Introduction} % Main chapter title

\label{Chapter1} % For referencing the chapter elsewhere, use \ref{Chapter1} 

%----------------------------------------------------------------------------------------

% Define some commands to keep the formatting separated from the content 
\newcommand{\keyword}[1]{\textbf{#1}}
\newcommand{\tabhead}[1]{\textbf{#1}}
\newcommand{\code}[1]{\texttt{#1}}
\newcommand{\file}[1]{\texttt{\bfseries#1}}
\newcommand{\option}[1]{\texttt{\itshape#1}}

%----------------------------------------------------------------------------------------

\section{Research Overview}
Speech, it is the tool that facilitates us with the ability to speak, orally communicate, and express oneself emotions through sound. Although animals have their own way of communicating through various sounds and inflections, Speech is often referred to as a form of human communication. This is because over time humans have gained the ability to control their voices by selective modification of various sound source spectrum to produce perceptible structures, which are used to convey different linguistic sounds (\cite{Speech}). Speech is more of a mechanical act but voice is what gives it a personal touch and is responsible for identification of an individual. When the air-stream leave the lungs, reaching the larynx a sound is generated due to the vibrations in the vocal folders present in them which is then passed on the pharynx where the voice patterns are shaped and then the speech is generated through the controlled movement of the the oral cavity (\cite{doi:10.1121/1.4964509}). The difference in the shape of the vocal cord and the vocal tract of every individual is responsible for the uniqueness in every persons voice. And this is the reason that voice can be used as a unique signature for ones identification. Us humans with all our cognitive abilities can identify the person and whether or not someone is trying to mimic a voice of other person in a fairly easy, simultaneous and efficient manner, even in the primitive stages in our life. But to automate such tasks and exploit the potential that voice and speech holds is a rather interesting yet perplexing process on it own.    

Given a single speech signal one can manually do the task of identifying the speaker in it. But if the task is meant to be commercialized, scaled globally it demands its own research field known as Automatic Speaker Verification (ASV) technology. This technology is focused towards extracting the speaker-specific information from the speech signal for verification of the speakers identity (\cite{jp1997speakerrecognition})(\cite{ae1976speakerverification}). But with the demand of ASV systems globally to be used as bio-metrics has raised concerns towards its reliability under unforeseen circumstances where the adversary might try to deceive the ASV system by claiming as another user. The claim by the adversary can be done either by mimicry or using technology to perform different kinds of spoofing attacks. The spoofing attacks(Presentation attacks) are mainly of three types: replay attacks, speech synthesis (SS), voice conversion (VC). These attacks are done with the ill intention of performing fraudulent activities like identity theft and forgery for selfish reasons. The technology used in these spoofing attacks deserves a separate research field of their own.

The core purpose of the ASV systems is to facilitate speaker identification and bio-metrics but these spoofing attacks take advantage of the incapability of ASVs to identify the artifacts caused by SS and VC systems trying to present adversary as original user. The exploitation of this loop hole by the adversary makes the ASV system less reliable and hinders its commercialization. This lead to the origination of another research field to detect spoofing attacks and develop systems known as anti-spoofing measures or countermeasures (CM) (\cite{z2014countermeasuresurvey}). These stand-alone CM systems are used before or in parallel with the ASV systems to eliminate the samples which seem fraudulent before-hand and let the ASV do what it is supposed to do without hindrances from any of the spoofing attacks. This kind of setup results in system that is more robust, reliable and ready to be scaled for voice bio-metric based tasks. 

In order to research and develop robust anti-spoofing methods the  \textit{Automatic Speaker Verification Spoofing and Countermeasure Challenge} (ASVSpoof) was designed. The latest spoofing challenge was ASVSpoof2019 in which the data and protocols were based on the VCTK corpus that encompassed two partitions for the assessment of Logical access (LA) and Physical access (PA) scenarios. From our earlier discussion speech synthesis and voice conversion are considered as attacks under the LA scenario. On the other hand replay attacks are considered in PA scenario. A complete description of the data set and the evaluation scenarios is available in the ASVspoof2019 evaluation plan~\cite{eval2019plan}. To develop a better understanding of how various spoofing techniques and countermeasure systems have developed over time few chapters are dedicated to give a brief overview about them.

%----------------------------------------------------------------------------------------

\section{Motivation for Detection of Doctored Speech}

The research on development of countermeasures for speech spoof detection originated to compensate for the incapability of the ASV system to capture the artifacts generated by the spoofing attacks. To get a better understanding of this subject it is better to have a brief overview of the concepts that builds up the premise for it, that is, ASV systems and what makes them more prone to spoofing attacks.

\subsection{Automatic Speaker Verification}

The concept of Automatic Speaker Verification (ASV) can be considered as a sub-task of Automatic speaker recognition (ASpR). Where the underlying proposition for speaker recognition is that of the difference between the pitch, loudness and timber of a person. Apart from the nature of the voice a few other factors like the style of speaking, which can originate from how tongue, lips and jaw are controlled, can help distinguish one person from another. The working of the ASpR system can be broken down into two phases: Enrollment and verification. In enrollment phase the speech sample of the person is taken and voice signature or print is formed by extracting specific features that represent the nature of the voice precisely. Once the voice signature is formed it is stored as one of many other voice signatures of other people already enrolled. And in the verification phase a speech sample is taken in order to verify the identity of the person (\cite{inproceedings}).

The ASpR system can be divided into two categories: text-independent and text-dependent. If it is mandatory to have same text in both the speech samples used for enrollment and verification then the ASpR system used is a text-dependent one. In such a system the text prompts can be common to all the speakers or can have possibility of having a password based multi-factor authentication system that uses both the nature of the voice as well as the password as a way to authenticate the user. In case of text-independent ASpR systems very less or negligible cooperation is needed by the speaker. The enrollment and verification phases can have speech samples with different text in them. Here the authentication or verification of the person happens only extracting features that capture the nature of the voice.

Finally, the architecture of ASpR system can be modified for two commercial applications: Automatic Speaker Identification (ASI) and Automatic Speaker Verification (ASV). ASI is the task of finding the identity of the speaker from a set of already enrolled voices. It determines the speaker in the given speech signal and it can be better understood if considered as a multi-classification problem where the classes are the set of enrolled speakers and the system has to determine the class of the input sample. In the case of ASV the task of accepting or rejecting the person claiming to be the actual speaker. It is more of a one-to-one matching task where the persons voice is matched with the voice signature of the enrolled speaker. To understand this better this scenario can be considered as binary classification problem where the classes are true speaker or imposter and the system has to determine the input sample belongs to which class. 

Before the introduction of End-to-End models the process of ASpR whether ASI or ASV can be outlined in an orderly manner using four main modules: (i) Preprocessing (ii) Feature Extraction (iii) Modeling (iv) Matching/Decision Logic. The preprocessing needed in the case of ASpR systems are mainly speech enhancement techniques such as denoising and channel compensation as filtering effects from the medium of the signal can also cause distortion. \cite{4769239} proposes a technique to subtract a suitable amount of noise from each frequency bin to prevent any distortion of the speech. And significant performance improvement over the spectral subtraction method was based on fast noise estimation proposed by \cite{Zhiming}. \cite{1660165} provides a MAP based channel compensation technique for their ASpR system. After enhancing the signal speaker dependent features are extracted which have large inter-speaker and small intra-speaker variability. This can be done on two levels using low level (\cite{Prahallad}), (\cite{Chakroborty}) and high level (\cite{4291592}),(\cite{879790}) feature extraction techniques. The modeling part can have three different perspectives, that is, speaker model generation (\cite{Aronowitz2005EfficientSI}), imposter modeling (\cite{kim2006impostor}) and cipher/Encryption (\cite{Enayah}. Then comes the Decision Logic/Matching where the main scope of improvement or experimentation is done in score normalization (\cite{JAIN20052270}), (\cite{Ramos}. With recent advances in deep learning a wide range of changes appeared in the architecture of the ASpR replacing the individual blocks with deep neural networks (\cite{6853887}),(\cite{zeinali}). Various schemes such as d-vector (averaged activation from the last layer of the DNN) and b-vector(using basic binary operation to form a high dimensional vector with augmentation) were used to replace individual feature extraction and binary classification blocks respectively (\cite{6854363}) (\cite{6853880}). With the entry of end-to-end systems which takes raw audio as input and give log probabilities as output have completely replaced the need for individual specific purposed block with CNN blocks that are arranged back to back to give results that by far exceeds the performance of the traditional models. To further improve on the convergence rate, performance and reducing the number of trainable parameters \cite{ravanelli2018speaker} introduced an ASpR end-to-end model Sincnet that replaces the normal CNN layer with a sinc function based convolution layer that encourages to capture even more meaningful features than the low-level speech features captured by the CNN filters. 

\subsection{Spoofing in ASV Systems}

ASV system are more fast and efficient as compared to ASI systems. The reason being that in ASV system the single comparison of the input signal with the reference pattern makes it a more tractable problem as compared to the huge number of comparisons that are needed in ASI systems. This feature of the ASV can has its own advantages and disadvantages. Its advantage being that its less complex nature makes it more scalable for commercial applications but this same reason leads to its downfall that it is more vulnerable to spoofing attacks. The commercial applications of ASV are related towards bio-metric operation, log-in credentials, authorizing banking transactions and many more. Which makes the issue of dealing with spoofing in ASV even more of an emergency then it was ever before.

If we were to examine the use of ASV systems, its main purpose is to verify the speaker signal by matching the signal with the target pattern and nothing else. With the recent advances mostly all the ASV systems are evaluated using casual attacks that require zero-effort (also called zero-effort attacks). These casual zero-effort attacks can range from a person trying to mimic the voice of target speaker to recording the voice of the target speaker (in an uncontrolled environment) and playing the recording in-front of the ASV system. The current ASV systems are robust to such attacks and tend to achieve high accuracy (low EER) when such attacks are thrown at them.

With the recent advances in technology, the stress test done on ASV systems using zero-effort attacks and evaluating the robustness of the system on the basis of those attacks is no where near the real world scenarios. \cite{7029420} in their paper carried out experiments to asses the threat of spoofing attacks of varied degrees on ASV systems (a detailed explanation of the spoofing attacks will be given in the next few chapters) and it was shown that ASV system when tested with the attacks using current technology will incur a huge dip in performance. Thus proper proactive stress tests must be designed to evaluate the performance of the system. And most importantly it is very necessary to spread awareness regarding such malicious attacks that the ASV system is vulnerable to. One such initiative is the ASVspoof challenge which will be described in the next section.

%----------------------------------------------------------------------------------------

\section{ASVspoof}

To spread awareness about presentation attacks and to conduct research in a systematic manner \textit{Automatic Speaker Verification Spoofing and Countermeasure Challenge} (ASVspoof) was designed. This challenge was originated from the special sessions about spoofing and countermeasures for ASV during the INTERSPEECH 2013 and has been conducted every alternate year since then. The vision of this challenge was to bring awareness about spoofing, build a competitive environment to fabricate self-contained anti-spoofing countermeasures in parallel to the ASV system and to congregate a body to collect and distribute databases that are evaluated on the basis of standard protocols and metrics (\cite{z2017cm}). The latest spoofing challenge was ASVspoof2019 in which the data and protocols were based on VCTK corpus that encompassed two partitions for the assessment of Logical access(LA) and Physical access(PA) scenarios. With full descriptions available in the ASVspoof2019 evaluation plan (\cite{eval2019plan}). To get a better understanding of the datasets used for the experiments in the thesis a brief comparison between the ASVspoof challenges is given in the subsection below .

\subsection{Comparison between ASVspoof challenges}

The first edition was about ASV spoofing was held in 2013 and was only focused on spreading awareness about the security threat that spoofing attacks can cause for ASV systems, ASVspoof 2015 was the first ever challenge to held on this topic. It provided a playground that allowed the development and comparison of various countermeasures on a common dataset with well defined protocols and metrics. The task of this competition was to develop a stand-alone spoofing countermeasure that can distinguish between spoofed and human speech. The 2015 challenge was solely based on the development of countermeasures that could on detect TTS and VC spoofing attacks. While the next ASVspoof challenge held in 2017 was only focused towards developing countermeasures that could detect only replay based spoofing attacks. 

Although the 2017 challenge was a success the dataset had some major flaws like uncontrolled environment during the creation of replay attacks which made it impossible to perform proper analysis on the given dataset. Due to this a new version of the dataset was published with controlled acoustic environments and meaningful replay characteristics that allowed improved and more general analysis for the replay attacks. while both of these challenges showed that there is lot of scope improvement in distinguishing spoofed and human speech. The ASVspoof 2019 challenge came along with its new and improved dataset. It was the first challenge that focused on all three types of spoofing attacks: TTS, VC and replay attacks. Although the attacks were build from the same VCTK dataset used in previous challenges but latest models and SOTA VC and TTS technologies were used to develop the attacks. The new dataset managed curated and recorded signals in a carefully controlled environment that gave even more revealing analysis than before (\cite{wang2019asvspoof}).   

\subsection{Database Details} \label{Database Details}

\subsubsection{Logical Access}

The ASVspoof database was fully Logical Access based and was generated according to a diverse mix of 10 different speech synthesis and voice conversion systems. The ASVspoof 2015 database was based on corpus called ``SAS corpus’’,further additional process such as removing broken files, trimming some silence frames, etc. for convenience. 
The LA database for ASVspoof 2019 contains bonafide speech and spoofed speech data generated using 17 different TTS and VC systems.Data used for the training of TTS and VC systems also comes from the VCTK database but there is no overlap with the data contained in the 2019  database. Six of these systems are designated as known attacks,  with the other 11 being designated as unknown attacks. The training and development sets contain known attacks only whereas the evaluation set contains 2 known and 11 unknown spoofing attacks.  Among the 6 known attacks there are 2 VC (\cite{VC2006transformation}) systems and 4 TTS(\cite{Morise2016WORLDAV}),(\cite{oord2016wavenet}), (\cite{Tanaka2018SynthetictoNaturalSW}) systems. 

\subsubsection{Physical Access}

The bonafide data and spoofed data contained in the PA database are generated according to a simulation of their presentation to the microphone of an ASV system within a reverberating acoustic environment. Replayed speech is assumed first to be captured with a recording device before being replayed using a non-linear replay device.Training and development data is created according to 27 different acoustic and 9 different replay configurations. Acoustic configurations comprise an exhaustive combination of 3 categories of  room  sizes,  3  categories  of  reverberation  and  3  categories of speaker/talker2-to-ASV microphone distances.  Replay configurations comprise 3 categories of attacker-to-talker recording distances, and 3 categories of loudspeaker quality.

ASVspoof challenges are focused on promoting awareness and fostering solution to spoofing attacks generated from state-of-the-art TTS, VC and replay technologies\cite{evans2013cm}\cite{z2017cm}. These challenges have recently adopted an evaluation metric to asses impacts of standalone anti-spoofing system to a fixed ASV system. This new primary metric is called t-DCF \cite{Kinnunen2018tDCFAD} and EER is now considered as the secondary metric.The following presents a summary of the specific characteristics of the performance measures and baselines. 

\subsection{Performance measures and baselines}
The recent challenge ASVspoof2019 emphasizes the assesment of tandem systems consisting of both a spoofing countermeasure(CM) and an ASV system(provided by the organiser). The performance of them combined systems is evaluated via the minimum normalized tandem detection cost function of the form:
\begin{equation}\mathrm{t}-\mathrm{DCF}(s)=min_{s}\{\beta P_{\mathrm{miss}}^{\mathrm{cm}}(s)+ P_{\mathrm{fa}}^{\mathrm{cm}}(s)\}\end{equation}
where B depends on application parameters(priors, costs) and ASV performance(false accept and false reject rates), while Pmiss(s) and Pfa(s) are the CM false reject and CM false accept rates at threshold s. The minimum in (1) is taken over all thresholds on given data(development or evaluation) with a known key , corresponding to oracle caliberation. While the challenge ranking are based on pooled performance in either scenario(LA or PA), results are also presented when decomposed by attack. In this case, B depends on the effectiveness of each attack. In particular, with everything else being constant, B is inversely being proportional to the ASV false accept rate for a specific attack: the penalty when a CM false rejects bonafide sample is heigher in case of less effective attacks. Likewise, the relative penalty when a CM false accepts spoofs is heigher for more effective attacks. The EER serves as a second metric. The EER corresponds to a CM operating point with equal false reject and false accept rates. And was the primary metric for previous editions of ASVspoof. Without a link to the impact of CMs upon the reliability of an ASV system, the EER may be more appropriate as a metric for fake audio detection, that is where there is no ASV system.  

ASVspoof2019 adopted two CM baseline systems. They use a common Gaussian mixutre model(GMM) back-end classifier with either constant Q cepstral coefficient(CQCC)\cite{todisco2017cqcc} features or linear frequency cepstral coefficient(LFCC)\cite{sahidullah2015ssdfeatures} features.\cite{Todisco2019ASVspoof2F}

%----------------------------------------------------------------------------------------

\section{Contributions}

This thesis is written to put forward not only the countermeasure systems developed during the tenure of this work but also in the best interest of acting as an up-to-date resource stating the state-of-the-art SS and VC technologies being used as spoofing attacks and countermeasure or anti-spoofing systems being developed to tackle these attacks till date. The entirety of the work in this thesis is done to facilitate the robustness and reliability of the ASV system. Every countermeasure covered in this work is so developed to extract distinctive features that enhance the difference between natural and spoofed speech so as to eliminate the the spoofed samples before it reaches the ASV system for speaker verification. It is to be noted that the extracted features are independent of the speech content of the signal and only focuses on the artifacts generated by the spoofing attacks. And the metric used to evaluate each and every countermeasure system in this work is exactly the same as used in ASVspoof challenge (\cite{z2017cm}), (\cite{eval2019plan}) , that is, tandem - Detection Cost Function (t-DCF) (\cite{kinnunen2018tdcf}) and Equal Error Rate (EER). To develop a better understanding of how various spoofing techniques and countermeasure systems have developed over time few chapters are dedicated to give a brief overview about them. The brief description of the contributions of this work are as follows:

\begin{itemize}
    
    \item An overview of the trends in non-proactive and proactive spoofing attacks 
    
    % \item A fair comparison between traditional handcrafted features and end-to-end countermeasure system
    \item Traditional handcrafted feature and end-to-end countermeasure system are trained and evaluated on modern spoofing attacks in order to perform a fair comparative analysis on them.
    
    \item Initial results on the ASVspoof2019 dataset on Wavelet based end-to-end architectures. These architectures are based on the Wavelet Scattering Transform and the discretized Continuous Wavelet Transform as feature extractors.
    \item Improvements in performance on the Sincnet architecture baseline~\cite{zeinali2019detecting} when results are fused from the said Wavelet based architectures with the outputs from the Sincnet architecture.
    \item Progress towards Parametric learn-able filters using wavelet based parametric filters in order to get insight about the prominent frequency bands responsible for distinguishing spoof and human samples
    
\end{itemize}

%----------------------------------------------------------------------------------------

\section{Thesis Organization}

The organisation of the thesis is discussed in detail below:
\begin{itemize}
    \item \textbf{Chapter 2} discusses the trends in the development of voice conversion and Text-To-Speech systems and the systems used in ASVspoof 2019 database. And give an over view about non proactive and proactive spoofing attacks.   
    
    \item \textbf{Chapter 3} gives an overview about the methodologies used over the span of ASVspoof challenge to develop countermeasures for spoof detection. The topics discussed will be related to the trends in preprocessing, feature extraction and various architectures used from classification.
    
    \item \textbf{Chapter 4} discusses about the experiments carried out using traditional handcrafted countermeasure systems(separate feature extractor and classifier) and test whether they are reliable enough against spoofing samples created by SOTA spoofing models. 
    \item \textbf{Chapter 5} introduces the concept of Continuous wavelet transform (CWT) and wavelet scattering transform (WST) and discusses about the proposed end-to-end architectures using CWT and WST as feature extractors to distinguish between spoof and bonafide samples. 
    \item \textbf{Chapter 6} gives an insight about the major frequency bands and scales responsible for spoof detection. This is done using wavelet based parametric learn-able filters.
\end{itemize}

It is worth mentioning that \textbf{Chapter 2} and \textbf{Chapter 3} give a brief description regarding the progress in spoofing attacks and Countermeasure systems, respectively. Furthermore, the rest of the chapters are self-sufficient, and each one of them builds a case for the subsequent chapter.

% Chapter 2

\chapter{Spoofing Systems} % Main chapter title

\label{Chapter2} % For referencing the chapter elsewhere, use \ref{Chapter1} 

%----------------------------------------------------------------------------------------

% Define some commands to keep the formatting separated from the content 
% \newcommand{\keyword}[2]{\textbf{#1}}
% \newcommand{\tabhead}[2]{\textbf{#1}}
% \newcommand{\code}[2]{\texttt{#1}}
% \newcommand{\file}[2]{\texttt{\bfseries#1}}
% \newcommand{\option}[2]{\texttt{\itshape#1}}

%----------------------------------------------------------------------------------------

\section{Introduction}
      
ASV system is considered as a matured technology. It has its applications in forensics, surveillance systems, access control and many more. But ASV systems which lack security measures are vulnerable to spoofing attacks. The attacks tend to exploit the weak points of the ASV systems at two main levels: Sensor level and software level. The replay attacks are used to fool the ASV at sensor level by playing recorded sample of the target speaker. While the TTS and VC systems are used to attacks the ASV at software level by feeding in synthetic or voice converted samples that match the voice pattern of the target speaker. In order to build spoofing systems it is important to have an appropriate knowledge of these spoofing systems. Another perspective to categorize the spoofing systems is through intention of the development of the spoofing attacks: non-proactive and proactive attacks. This perspective about spoofing attacks will be discussed in the \ref{Non-Proactive and proactive attacks} section of this chapter.

One way to measure the progress of the spoofing systems is to have a brief idea about the spoofing attacks used in the ASVspoof challenge. The reason that this might be a fair way to get an idea about the spoofing systems is due to the fact that the main motive of the ASVspoof challenge initiative was to gain awareness and build a standard method for competitive evaluation of countermeasures developed in order to protect the ASV systems against the so called moving target of VC, TTS and replay technologies. The section \ref{History of spoofing systems} will discuss about the underlying technology for spoofing attacks used in ASVspoof 2015 dataset. And the spoofing attacks used in ASVspoof 2019 will be discussed in the section \ref{Current trends in VC and TTS} of this chapter.

%----------------------------------------------------------------------------------------

\section{Types of spoofing attacks}

As already discussed in section \ref{Database Details} the spoofing systems can attack the ASV at two levels, Senor level that is the physical access attacks and Software level that is the Logical access attacks. For the scope of this thesis we will only discuss Logical access attacks. The Logical access attacks are of two types: Text-to-speech and Voice conversion based systems. A brief description about them are as follows:

\subsection{Text-to-speech systems}

As the name suggests the main goal of TTS is to generate intelligible human like speech from text. The conventional method of doing this can be summarized as follows : text analysis, text normalization, text processing, grapheme-to-phoneme conversion and speech synthesis. Text analysis processes the input and divided them into small tokens like words and sentences. Text normalization is the removal of punctuation's and converting the upper case letters to lower case. The process of assigning phonetic transcription to word is called grapheme to phoneme conversion. Finally speech is synthesised from the phonetic transcriptions.  

\subsection{Voice Conversion Systems} \label{Overview of Voice Conversion}

The main objective of voice conversion systems is to use the natural voice of the attacker and convert it into the target voice. The voice conversion system directly work on speech unlike the TTS which require text. Modules like vocoders that generate the speech are common in both. The application of voice conversion lies in many areas such as (a) restoration of helium speech (b) mimicking voice for creating spoofing attacks and (c) preprocessing speech for recognition to make the speech recognition system more robust and reliable.

\section{History of Spoofing Systems} \label{History of spoofing systems}

This section gives a brief overview of the logical access spoofing systems used to make the ASVspoof 2015 dataset. Here are some of the TTS and VC systems used in ASVspoof 2015 dataset: 
\begin{itemize}
    \item Simple voice conversion technique which only modifies the first coefficient of Mel-Cepstral coefficients. 
    \item A naive frame-selection based voice conversion.
    \item HMM-based speaker-adapted speech synthesis
    \item Voice conversion system based on maximum likelihood estimation of spectral parameter trajectory
    \item MaryTTS unit selection based text to speech system
\end{itemize}

% And the details about the setup used for recording the voice of target speaker for the development of replay attacks in ASVspoof 2017 database is given in the section \ref{Overview of Replay}. 

\subsection{Unit selection based text to speech system} \label{unit_selection}

The most common speech synthesis system before the arrival statistical parametric speech generation was the unit selection technique which was based on the concept of picking up part of word or sub-words from a database of large vocabulary of sub-words and stitch them together to form an utterance.But this technique with its advantage of high quality speech due to waveform concatenation had several downfall as follows:
\begin{itemize}
    \item Discontinuity present in the output sample due to stitching of various small utterances.
    \item Hit or miss during selecting the part of words.
    \item Huge memory requirement due to large vocabulary database size.
    \item Lack of customisation in the utterances selected, that is to portray all possible intonations and prosodies the database size will increase exponentially. 
\end{itemize} 

The concept of unit selection and how to minimize its disadvantages is better explained by the following example:

% \begin{equation} \label{equationmaryTTS}
% \operatorname{total} \operatorname{cost}\left(u_{i}\right)=W_{1}^{\mathrm{T}} *\left(\begin{array}{c}
% \operatorname{target} \operatorname{cost}\left(u_{i}\right) \\
% j \operatorname{oin} \operatorname{Cost}\left(u_{i}, u_{i-1}\right) \\
% \operatorname{scost}\left(u_{i}\right)
% \end{array}\right)\end{equation}
\begin{equation}
\text { totalcost }\left(u_{i}\right)=W_{1}^{\mathrm{T}} *\left(\begin{array}{c}
\mathrm{targetcost}\left(u_{i}\right) \\
j o i n \mathrm{Cost}\left(u_{i}, u_{i-1}\right) \\
\mathrm{scost}\left(u_{i}\right)
\end{array}\right)
\end{equation}

The unit selection generally consist of two types of costs, One for defining how well the sub-word candidate matches the target unit. This cost is called the $targetCost$. Another cost is the $joinCost$ that defines how well the selected unit candidates can combine at the joins. Apart from this \cite{charfuelan2013mary} has proposed a cost function to select the best candidate with the main objective of minimizing the total cost given in \ref{equationmaryTTS} where $u_i$ is the $i^th$ unit candidate; $c$ is the cost vector; and $w$ is the weight vector for these features. In addition to these
two costs, each unit candidate is associated with a precomputed
$sCost$. The statistical cost ($sCost$) is calculated as follows:

\begin{itemize}
    \item Segment labels are estimated on the basis of the recorded speech and transcriptions from prompts
    \item The labels, transcriptions and the recording are used to create the HMM based voices. The author used HMM-based voice with a stable, not too expressive, narrative style to create a neutral voice from the audiobook data.
    \item spectral parameters,that is, Mel generalised cepstrum (mgc) features are extracted from the recordings
    \item HMM-based voices with the transcriptions provided are used for generating spectral features in the MARY TTS framework. For performing the alignment labels (phone durations) generated for each sentence are also kept.
    \item Forced alignment between the extracted spectral features and generated Mel cepstrum parameters are done using labels and dynamic time warping (DTW). Then finally calculate an optimum path sCost measure. Mahalanobis distance between the extracted spectral feature vector and generated mel cepstrum parameters vector is used to find the optimal path which is divided by the number of frames in the recorded segment and in the generated segment, that is the sCost. Here Mahalanobis distance is the criterion for finding the optimim path.
\end{itemize}

\subsection{HMM-based speaker-adapted speech synthesis}

\begin{figure}[h]
\centering
\caption{Overview of a typical HMM-based speech synthesis
system. \cite{zen2007hmm}}
\includegraphics[width=12cm]{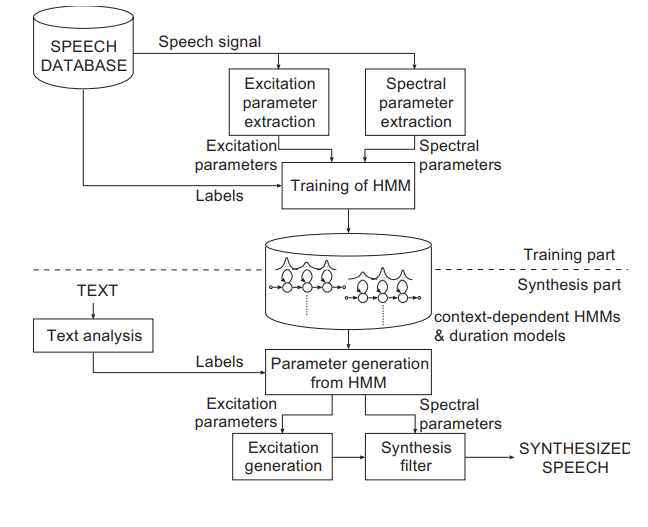}
\label{fig:HMMsynthesis}
\end{figure}

To cover up for all the dis-advantages of the unit selection statistical parametric speech synthesis techniques were preferred for speech synthesis. This concept is explained using the examples of HMM-based speaker-adapted speech synthesis (\cite{zen2007hmm}). The process of speech synthesis using HMM's can be divided into two phases: training and synthesis phase. The training phase is similar to that of the speech recognition systems. The only difference between the training phase of speech synthesis and speech recognition is where the context dependent HMM's are modeled using the spectrum parameters which are the MFCC and its delta features, and excitation parameters(log fundamental frequencies ($logF_0$)) that are extracted from the speech database. The context dependent HMM's use the phonetic, linguistic, and prosodic contexts to model the sequence. Multi-space probability distribution(MSD) and probability density function(PSD) are used to model parameter sequence that are multi-dimensional ($logF_0$) with unvoiced regions and to capture temporal structure of speech respectively.

The inverse of speech recognition is the process of speech synthesis. First, context-dependent label sequence is made from an arbitrarily given text that is to be synthesized, and then we concatenate the context dependent HMM's according to the sequence labels to form an utterance HMM. Second, the state duration PDF's are used to determine the state duration's of the utterance HMM. Third, to generates the sequence of spectral and excitation parameters the speech parameter generation algorithm is used. Finally, using the corresponding speech synthesis filter(MLSA) a speech waveform is
synthesized directly from the generated spectral and excitation
parameters. To get an idea of the speech generation algorithm used refer to \cite{zen2007hmm} And the while sequence of training and synthesis is depicted in the following figure \ref{fig:HMMsynthesis}.

Some of the other TTS based systems used in ASVspoof 2015 were also similar to HMM-based speech synthesis with other modifications like speaker adaptation and constrained structural maximum a posteriori linear regression (SMAPLR) based algorithm (\cite{Yamagishi})

\section{Current trends in VC and TTS} \label{Current trends in VC and TTS}

The collection of VC and TTS based spoofing attacks used in the ASVspoof 2019 is a result of huge amount of effort with contributions from various institutions and labs. There are a total of 19 VC and TTS based spoofing models that combined together form the database. All of these 19 spoofing attacks are the latest SOTA level models that were released mostly during or after 2018. 6 of these spoofing systems were known spoofing attacks, that is they were present in the training and development set of the database while all the other spoofing models were unknown attacks and were part of the evaluation set. To give the readers an understanding about the current SOTA systems of VC and TTS this section will give you a brief overview of few of the spoofing attacks used in the ASVspoof 2019 dataset. 

\begin{itemize}
    \item One of the attacks are based on neural-network (NN)-based TTS system.It uses a powerful neural waveform generator called WaveNet(\cite{oord2016wavenet}) . WaveNet vocoder can produce high-quality speech that fools CMs. Other attacks similar to this one are also present in the database but with different vocoders like WORLD vocoder or using the vocoder from merlin toolkit (\cite{wu16_ssw}).
    \item TTS systems based on concatenation of waveform, major advantage of using this is because it preserves the short-term acoustic features of natural speech (\cite{schroder:hal-00661061}).
    \item VAE based voice conversion system. This can be used because of its ease to build as compared to other traditional VC models and most importantly it does not require parallel speech data (\cite{inChin-ChengandHwang}).
    \item Voice conversion system based on transfer function which is known for increasing the false acceptance rate in the ASV system (\cite{Drissinproceedings}).
    \item GAN post filter based TTS system used to better mask the differences between natural and spoofed speech (\cite{tanaka2019wavecyclegan2}).
    \item A statistical parametric based TTS system which is real time and is based on Vocaine vocoder. Its main advantage is that it can run on mobile and reduce computational load (\cite{Zeninproceedings}).
    \item Transfer learning based end-to-end TTS called the Tacotron 2 (\cite{shen2018natural}), that uses knowledge from the field of ASV and applied to perform voice conversion.
    \item A NN-based system that does the job of both  VC and TTS system. This approch produces waveform using  a waveform modification method (\cite{Kobayashiarticle}, \cite{pmlr-v37-li15}).
    \item VC system that is based on the idea that was originally used by text-independent ASV systems. It is based on i-vector based on transfer learning method where  the pre-training is done to optimize the ASV system. The system is based on i-vector PLDA system that is then directly used to define the voice conversion function in a regression setting. MFCC features are used as input to generate speech from vocoder (\cite{Kinnuneninproceedings}).
\end{itemize}

%----------------------------------------------------------------------------------------

\section{Non-proactive and proactive attacks} \label{Non-Proactive and proactive attacks}

This section focuses on a broader issue about the attacks used in ASVspoof 2015 and 2019 database. Even though the ASVspoof community are using SOTA VC and TTS systems to build their database there is still one issue that does not make the countermeasures developed to protect the ASV systems up to the level that it can be used in real world scenarios. The TTS and VC systems main objective or goal is to model the human voice and this can be done by minimizing the difference between the spectral representation of the human voice from that of the TTS or VC model. These systems are designed to mimic the human voice but not to break the ASV system. The maximum they can do is to stress test the ASV system but not wreck it. Attacks posed by these kind of spoofing systems are called non-proactive spoofing attacks. And the attacks that are particularly designed to break the ASV system are called the proactive attacks.

In order to level up the game of security of the ASV system and make the ASV more robust it is not only important to develop countermeasures but also to think from attackers perspective because it will give a new perspective towards building the ASV system. It will give an insight about the weak links in the ASV system that the attacker can exploit. If an attacker were to attack an ASV system the most feasible part to attack to break the ASV was the functional module. But that is not possible as to do that the attacker needs to have some prior knowledge about the ASV's internal working. Another way to proactively attack the ASV system with spoofed samples that are specifically designed to break it. Such kind of spoofing attacks are called adversarial attacks. they are popular in image domain but not so much in speech. 

\begin{figure}[h]
\centering
\caption{non-proactive attacks and proactive attacks with black, white and grey box ASV's (\cite{das2020attackers})}
\includegraphics[width=12cm]{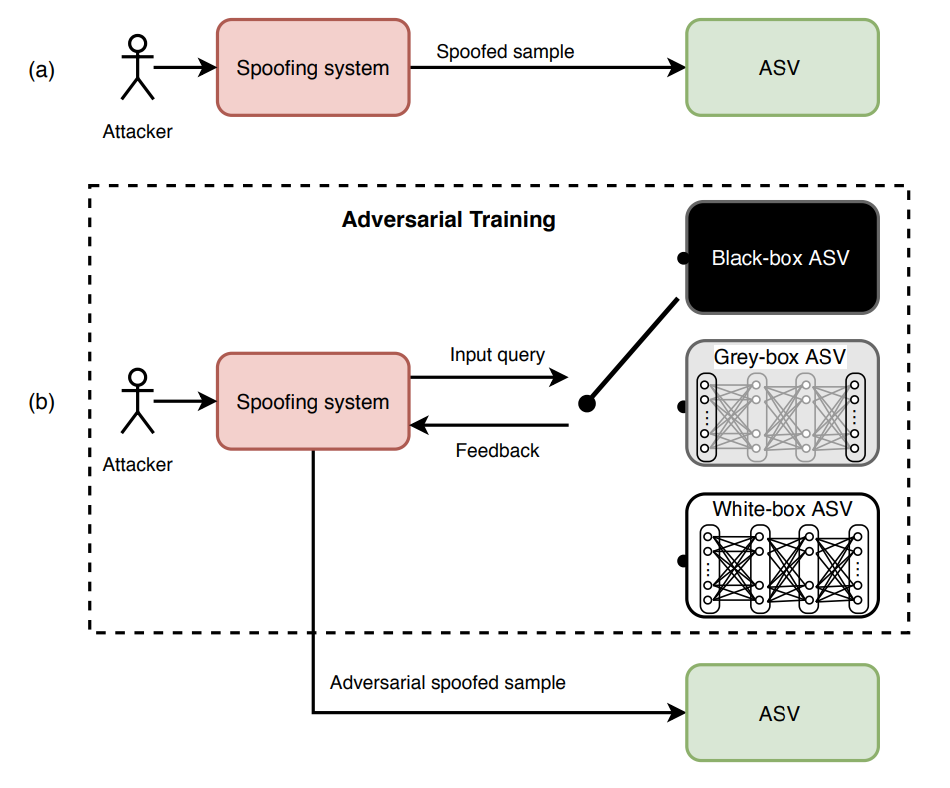}
\label{fig:attackerperspective}
\end{figure}

In figure \ref{fig:attackerperspective} The part a) is an illustration of the non-proactive attacks whereas the part b) describes the various scenarios that are possible when carrying out the proactive attacks/ adversarial attacks. In case of black-box ASV the attacker has zero knowledge about the internal working of the ASV and the only thing to generate the attack for the ASV are the output scores given by the ASV system. In case of grey-box ASV the attacker has a little more option to play with than before. It still does not have the full access to the statistical model used in the ASV but it has some idea about the internal working. For white-box ASV the attacker has full knowledge about the ASV and this means the security of the system is at great risk. The concept of building an artificial signal that is modified specifically break the attacked system is recent but not that uncommon. Such signals tend to have zero resemblance to human voice. Thus, this can be counted as one kind of adversarial attack but another kind which is not so common and will be more practical in the ASV spoofing scenario is the one where the signal is indistinguishable from human speech but is still capable of breaking the ASV system. An examples of such adversarial attacks are based on $fast gradient sign method$ which have done their experiments on black-box and white-box attacks with same ASV's (\cite{li2019adversarial}), another example is $FakeBob$ that use black-box attacks by adding artifacts to human speech similar ot the previous example but uses various ASV's to get a practical idea (\cite{chen2019real}).

% Chapter 3

\chapter{Countermeasure Systems} % Main chapter title

\label{Chapter3} % For referencing the chapter elsewhere, use \ref{Chapter1} 

%----------------------------------------------------------------------------------------

% Define some commands to keep the formatting separated from the content 
% \newcommand{\keyword}[2]{\textbf{#1}}
% \newcommand{\tabhead}[2]{\textbf{#1}}
% \newcommand{\code}[2]{\texttt{#1}}
% \newcommand{\file}[2]{\texttt{\bfseries#1}}
% \newcommand{\option}[2]{\texttt{\itshape#1}}

%----------------------------------------------------------------------------------------

\section{Introduction}
      
\begin{figure}[h]
\centering
\caption{Block diagram of a typical presentation attack detection system mostly used in ASVspoof 2015. (\cite{sahidullah2015ssdfeatures})}
\includegraphics[width=12cm]{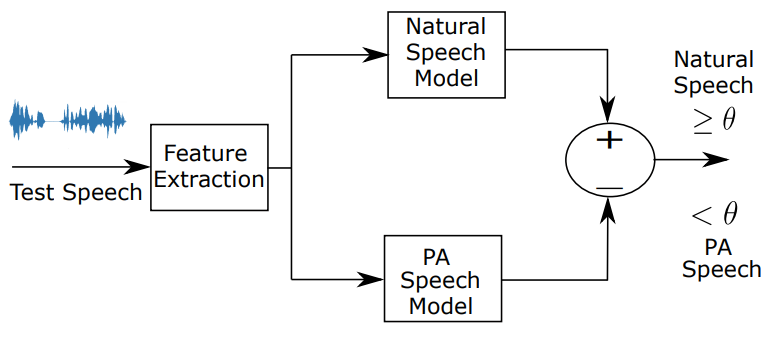}
\label{fig:oldarchitecture}
\end{figure}

In early years the selection of appropriate features for a given classification problem is an important task. But with the use of end-to-end deep learning and other similar techniques this classic boundary to think between a feature extractor and a classifier as separate component is getting increasingly blurred. But in the context of anti-spoofing the research on the feature extractor or front-end component still holds importance. This allows the utilization of prior knowledge of Digital signal processing and speech to shape the design of new discriminative features. For instance, past experience show that a speech generated using an approximate model such as TTS or VC systems cannot capture all the spectral complexities that a vocal tract system generates. Hence, there will be a significant difference between system-based features for natural and synthetic speech (\cite{sahidullah2015ssdfeatures}) (\cite{Wu2012DetectingCS}). Moreover, these spectral features are extracted and processed frame-wise while the human speech production is a continuous process. These initial findings sparked further research into developing advanced front-end features (\cite{sahidullah2015ssdfeatures}) (\cite{Wu2013DetectingCS}). To give an idea of the previously used architecture to detect spoofing attacks given in ASVspoof 2015 dataset are given in the Figure \ref{fig:oldarchitecture}

%----------------------------------------------------------------------------------------

\section{History of countermeasure Systems} \label{History of countermeasure systems}

The research work in anti-spoofing was divided into one of the following two categories: Front end Feature learning and Back end classifiers. The front end features are of many types, here are some to name a few: system-based features, source-based features, source-filter interaction features ,scattering cepstral coefficients, fundamental frequency variation features, short-term power spectrum features, short-term phase features and spectral features with long term processing. A list of features used in ASVspoof 2015 as to develop countermeasure systems is given in table \ref{fig:oldcountermeasure}.The first three features are based on type of information, That is the system-based features capture the perception mechanism occurring in the ear, the source-based features capture the mechanism of the vocal tract system, source-interaction features captures the interaction between the excitation source and the vocal tract system. All of these features tend to represent attributes of natural speech. Hence, any discrepancy from the characteristics of natural speech might be properly represented using these features. And the above feature sets are hand-crafted and consists of fixed sequence of standard signal digital processing operations. Al the features mentioned are either trying to model human speech or trying to model the perception model. So we can say that the features after the first three features belong to either one of the three features. To have a proper understanding of the front-end features the system-based features, source based features and source filter interaction features are explained as follows:

\begin{figure}[h]
\centering
\caption{Details about the front-end features used in ASVspoof 2015 and their respective EER's. (\cite{sahidullah2015ssdfeatures})}
\includegraphics[width=12cm]{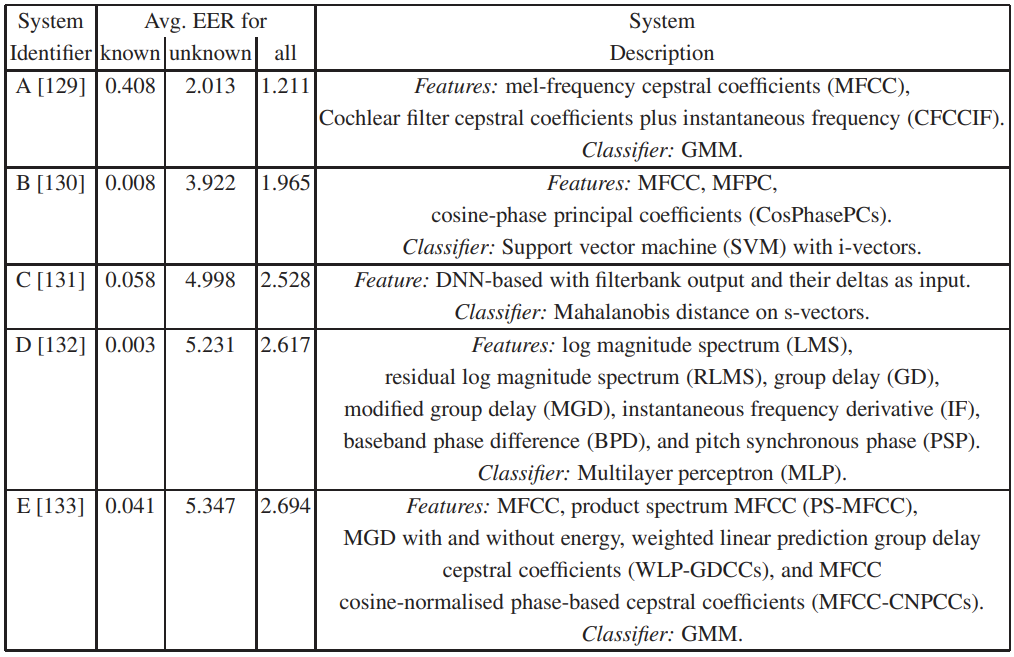}
\label{fig:oldcountermeasure}
\end{figure}

\subsection{System-based features} \label{system-based features}

The TTS or VC models for generating human like speech need to develop a mechanism to very closely approximate the model of the spectral representations that are perceived by the human ear. But however close the approximation models are to modelling the spectral representations they cannot exactly capture the patterns created or the structure of the vocal tract or this is what was true until deep learning models came into picture. So the spoofed samples used in ASVspoof 2015 dataset could be easily distinguished by the human ear. From the above discussion we can infer that in order to distinguish between spoofed and human samples one needs to build some sort of handcrafted features that can exploit and implant the perception mechanism of the human ear.

\begin{figure}[h]
  \centering
   \caption{Frequency response for a Mel filterbank with 20 filters and Frequency response for cochlea filterbank with 14 filters}
  \begin{minipage}[b]{0.5\textwidth}
    \includegraphics[width=\textwidth]{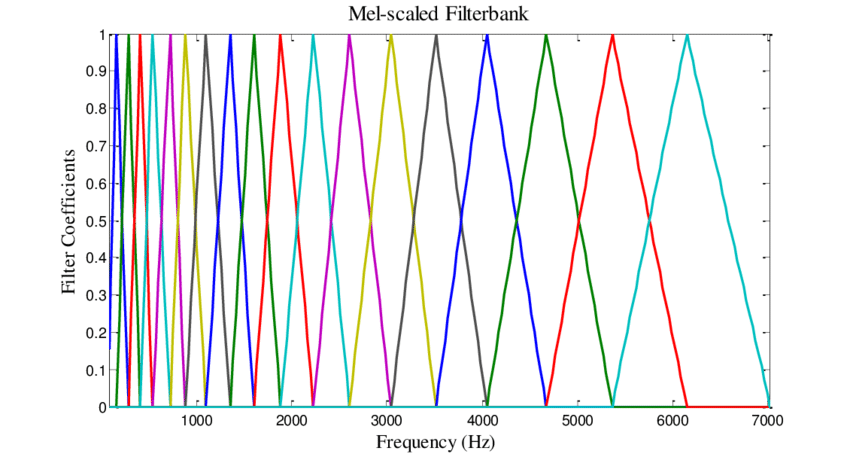}
  \end{minipage}
  \hfill
  \begin{minipage}[b]{0.4\textwidth}
    \includegraphics[width=\textwidth]{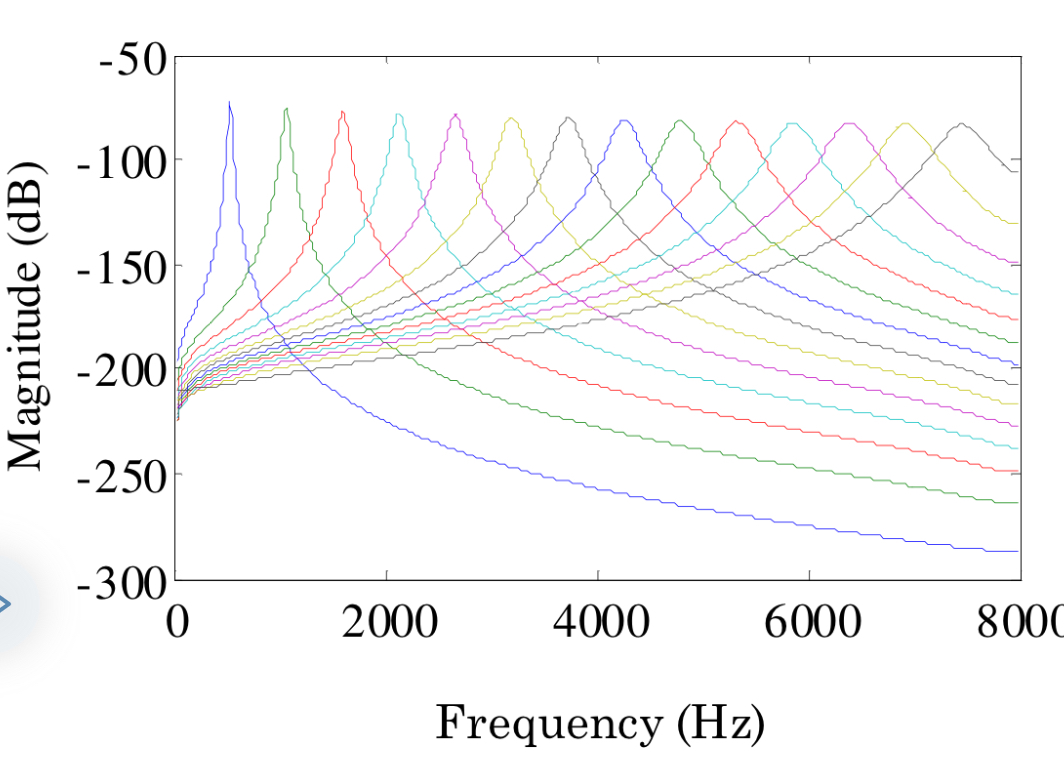}
  \end{minipage}
\label{fig:frequency response cochlea}
\end{figure}

If we study the mechanism of the human ear we observe that the speech sound reaches the inner part of the human ear where the cochlea is present and inside it  basilar membrane (BM) is present. This causes the BM to vibrate. These vibrations are not random in nature instead they vibrate at different regions for different tone of sound. This means we can consider the cochlea as a sub-band filterbank that can produce various frequency responses for different input frequencies. As you can see such phenomenon can be captured by certain signal processing abstractions. Such abstractions were brought to us by the MFCC features which have their center frequencies aligned according to the Mel scale and have symmetric nature triangular filterbank. These filters are compressed at low frequencies and well spaced at high frequencies. However the auditory filters also called the cochlea filters are asymmetric in nature which bring us to Cochlear Filter Cepstral Coefficients (CFCC) (\cite{7802564}) which use auditory transform (AT) (\cite{7459917}) that use sub-band filters that are designed to potray the hearing mechanism of the ear. The frequency response of both the mel filterbank and the cochlea filterbank is given in figure \ref{fig:frequency response cochlea} 

% \begin{figure}[h]

% \begin{subfig}
% \includegraphics[width=15cm]{Figures/22943.jpg} 
% \caption{CFCC extraction process}
% \label{fig:subim1}
% \end{subfig}
% \begin{subfigure}
% \includegraphics[width=15cm]{Figures/89046.jpg}
% \caption{MFCC extraction process}
% \label{fig:subim2}
% \end{subfigure}

% \caption{Block diagram for MFCC and CFCC extraction process}
% \label{fig:image2}

% \end{figure}
\begin{figure}
  \centering
  \begin{tabular}{@{}c@{}}
    \includegraphics[width=15cm]{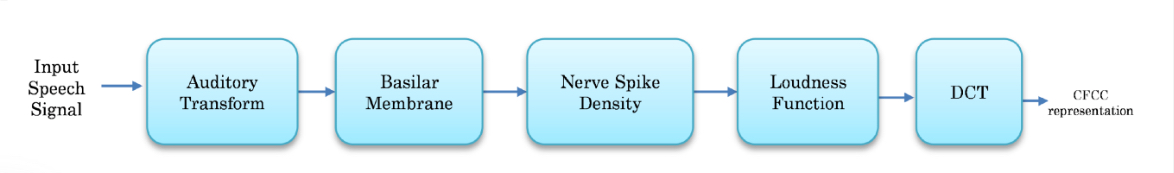}
    \\[\abovecaptionskip]
    \label{fig:subim1}
    \small (a) CFCC extraction process
  \end{tabular}

  \vspace{\floatsep}

  \begin{tabular}{@{}c@{}}
   \includegraphics[width=15cm]{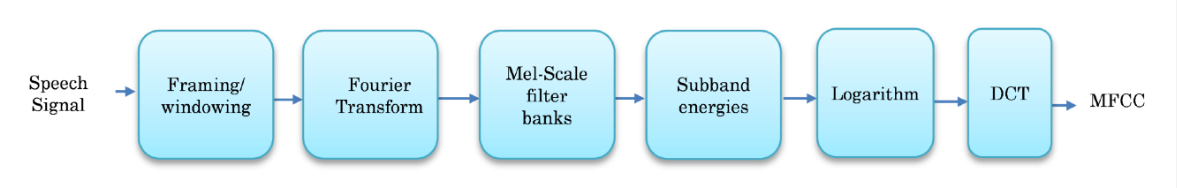}
   \\[\abovecaptionskip]
    \small (b) MFCC extraction process
    \label{fig:subim2}
  \end{tabular}

  \caption{Block diagram for MFCC and CFCC extraction process}\label{fig:image2}
\end{figure}

\subsection{Source-based features} \label{Source-based features}

As discussed in section \ref{system-based features} the TTS and VC in order to sound human like need to  model the human voice very closely but their approximation model cannot do so perfectly. The spoofing systems can also exploit this incapability of the spoofing systems to distinguish between spoof and human speech. In order to exploit this incapability of the spoofing systems one needs to build the handcrafted features that captures the dynamics of the vocal tract. The source based features are exactly supposed to do that. Fundamental frequency ($F_{0}$), strength of excitation ($SoE$), Linear prediction (LP) and non-linear prediction (NLP) are some of the source based features that were used in ASVspoof 2015 challenge to distinguish between spoof and human samples. 

$F_{0}$ and dynamic variations of $SoE$ that are derived from the speech and the excitation source. When speech is produced by the vocal tract, the vibration of the vocal fold affect the $F_{0}$ and amplitude of the speech. This variation of fundamental frequency and strength of excitation is absent in synthetic speech produced using the TTS or VC systems. And the motive for using NLP based countermeasures was due to the qualities of various NLP schemes of capturing the non-linear dependencies that natural speech have. Using source based features alone for detecting spoofing attacks are proven not be sufficient, and they do not provide comparable results to the system based features. But when score level fusion of the source based and system based features they do improve the performance as compared to the system based features alone. 

\subsection{Source-filter interaction features}
    
In section \ref{system-based features} and \ref{Source-based features} we discussed about handcrafted features that captures the properties of perception mechanism of ear and source excitation model of the vocal tract. But there is one more important feature that is responsible for the naturalness of the speech produced by the vocal tract, that is the non-linear interaction between the excitation source and the filter. 
This sort of interaction between the source and the filter is even more difficult to model by the text-to-speech or Voice conversion systems. 

\begin{figure}[h]
\centering
\caption{Block diagram of a typical presentation attack detection system mostly used in ASVspoof 2015. (\cite{7879263})}
\includegraphics[width=15cm]{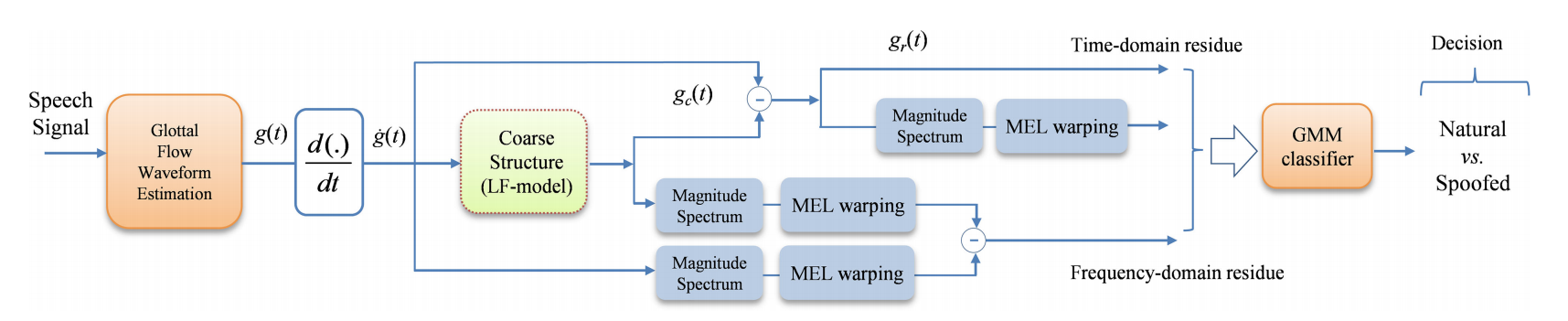}
\label{fig:SFinteraction}
\end{figure}

The work proposed in \cite{7879263}, the voice excitation
source (i.e., differenced glottal flow waveform, $\dot(g)(t)$) is estimated and Liljencrants–Fant model is used to model the voice excitation source, to get coarse structure, $g_{c}(t)$. Then difference between $\dot(g)(t)$ and $g_{c}(t)$ gives $g_{r}(t)$ which is known as the residue. This residue is known to capture the nonlinear S–F interaction. And the authors also noticed that during the time domain, if L2 normalization is applied to $g_{r}(t)$ in various stages of the glottis can also be considered as features. And in frequency domain, if the $g_{r}(t)$ is represented at Mel scale shows significant improvement over MFCC features alone. Residual energy features, Mel representation of the residual signal, and
Mel frequency cepstral coefficients (MFCC) when fused at score level gave an EER of 0.017\%, as compared to the 0.319\%
obtained with MFCC alone. Furthermore, same process of calculating of the residues was also done on spectrogram  of $\dot(g)(t)$ and $g_{c}(t)$ can also be used for the spoof detection task. The proposed approach for glottal source features for spoof detection is given in figure \ref{fig:SFinteraction}

%----------------------------------------------------------------------------------------

\section{Current trends in Countermeasures} \label{Current trends in countermeasures}

An alternative approach, seeing increased popularity across different machine learning problems, is to learn the feature extractor from given data by using deep learning techniques\cite{Tian2015verification}. This work uses deep neural network to generate bottleneck features for spoofing detection, that is the activation of the hidden layer with a relatively small number of nodes compared to the size of the other layers. The study in \cite{Qian2016deepfeatures} investigates various deep features based on deep learning techniques. General observation regarding front-end for speech spoof detection. The first refers to the use of dynamic coefficients\cite{sahidullah2015ssdfeatures}. The first and second derivatives of static coefficients are found important to achieve good spoof detection performance. This is not entirely surprising, since VC and TTS may fail to model the dynamic properties of the speech signals, introducing artifacts that help in spoof detection. The second refers to the use of speech activity detection. This could be a database dependent observation, since in ASVspoof2015 corpus keeping the silence regions was beneficial for discriminating between natural and synthetic speech. This is likely due to the fact that non-speech regions are usually replaced with noise during VC or TTS operation. Advances in back-end classifiers: In anti-spoofing classification problem, two main families of approaches have been adopted, namely generative and discriminative. Generative approaches include those of GMM-based classifiers\cite{Qian2016deepfeatures} and i-vector representations combined with SVMs. As for discriminative approaches, deep learning based techniques have become more popular. Finally, new deep learning end-to-end solutins are emerging\cite{Lai2019ASSERTAW}. In-fact in ASVspoof2019 results the top 7 in LA and top 6 in PA system used neural networks, which ensures that neural network is the correct way to go in anti-spoofing classification problem.
%----------------------------------------------------------------------------------------

% Chapter 4

% \chapter{Preliminary work: Wavelets vs MFCC and credibility of handcrafted traditional model in detecting latest spoofing attacks } % Main chapter title

\chapter{Preliminary work: Wavelets vs MFCC } % Main chapter title

\label{Chapter4} % For referencing the chapter elsewhere, use \ref{Chapter1} 

%----------------------------------------------------------------------------------------

% Define some commands to keep the formatting separated from the content 
% \newcommand{\keyword}[2]{\textbf{#1}}
% \newcommand{\tabhead}[2]{\textbf{#1}}
% \newcommand{\code}[2]{\texttt{#1}}
% \newcommand{\file}[2]{\texttt{\bfseries#1}}
% \newcommand{\option}[2]{\texttt{\itshape#1}}

%----------------------------------------------------------------------------------------

\section{Introduction}
      
The most popular handcrafted feature used for almost all speech related tasks from speech recognition to speaker recognition are the MFCC features (\cite{6838564})(\cite{6189918}). This can be due to the ability of the mel cepstral features to capture the source based qualities of the human speech. But with recent advancements in this field has led to the discovery of the utility of wavelets in speech. Despite the prior knowledge of the MFCC the multi-resolution analysis of the wavelets make them comparable to MFCC features. Such sort of analysis allows them to capture features and small artifacts that are useful in speaker recognition and speech enhancements (\cite{turner2015wavelet})(\cite{lee2003wavelet}). Due to such interesting set of researches presented for wavelet based features and MFCC motivated us to do a comparative study for them in the scope of spoof detection.     

This chapter acts as a basis for all the experiments that are presented in the future chapters of this thesis. The experiments performed in this chapter gives us an idea about the credibility of the handcrafted features in detecting latest spoofing attacks. Here the experiments are done on two spoofing databases: ASVspoof 2015 and ASVspoof 2019. The experiments done on the ASVspoof 2015 database will act as a justification for our inclination towards wavelet based features instead of MFCC features. The architecture used in our experiments will also act as bridge between using separate handcrafted features as feature extractor and GMM as classification model to using DNN as a substitute for both and even more blurs out the classic boundary between a separate feature extractor and classifier. The experiments on the ASVspoof 2019 database show the lack of credibility of the traditional handcrafted features and gives us all the more reason to progress towards using end-to-end deep neural networks and more recent techniques. 

The rest of the chapter is organized as follows. Section~\ref{MFCC} and \ref{MWPC} presents the brief description of the MFCC and MWPC features used in the experiments. Section~\ref{experiments} presents experimental details and results. Section~\ref{discussion} provides a discussion of the experimental results while Section~\ref{conclusion} concludes the chapter.

\section{Mel Frequency Cepstral Coefficients (MFCC)} \label{MFCC}

\begin{figure}[h]
    \centering
    \includegraphics[scale = 0.6]{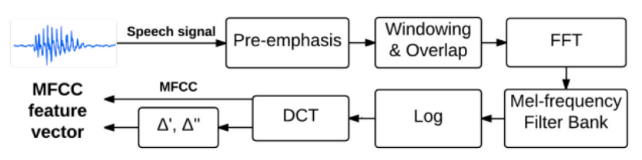}
    \caption{MFCC feature extraction pipeline(\cite{Alam2016features}).}
    \label{fig:MFCC_pipeline}
\end{figure}

The human voice is a by product of the opening and closing of the vocal folds present in the larynx due to the air pressure created by the lungs. And it is observed that the shape of the vocal tract can be manifested in the envelope of the short time power spectrum. The strong point of the MFCC features are that they can accurately capture this envelope. Here is MFCC implementation pipeline given in the figure \ref{fig:MFCC_pipeline}  and a brief description of the implementation of MFCC features used in our experiments:

\begin{itemize}
    \item \textbf{Pre-emphasize} is done to improve the signal to noise ratio, increase the high frequency energies as they usually have smaller magnitudes compared to lower frequency energies, avoid calculation difficulties in the later stage of the Discrete Fourier transform. Equation \ref{pre-emphasis}
    
    \begin{equation} \label{pre-emphasis}
    y(t)=x(t)-\alpha x(t-1)\end{equation}
    
    Here $y(t)$ is the resultant signal after pre-emphasis on $x$ where $t$ denotes every time sample. And $\alpha$ denotes the filter coefficient which is set to 0.97 for our experiments.
    
    \item \textbf{Framing} is done by splitting the signal into short enough time frames where the frequency content of the signal remains stationary. This step is done on the basis of an assumption that the frequency content of the signal does not change for a very short period of time. This step will make more sense while calculating DFT as it wont be able to accurately capture the frequency content of the signal if the time window is too large.
    
    \item \textbf{Windowing} is done to counter-react with effects of the assumption made by DFT. DFT assumes that the signal chunk we get after framing is periodic and infinitely repeating in nature. Due to this there are sharp transitions that create high frequency noises. So to avoid the sharp transitions hamming window function is multiplied to the framed signal to limit the value of the end to zero.
    
    \item \textbf{Discrete Fourier Transform (DFT) and Power Spectrum} is calculated over the framed and windowed signal. N-point Fast Fourier Transform can be used to calculate the DFT. And then power spectrum can be calculated using the formula given in equation \ref{power_spectrum}
    
    \begin{equation} \label{power_spectrum}
    P=\frac{\left|F F T\left(x_{i}\right)\right|^{2}}{N}\end{equation}
    
    Here value of $N = 512$ and $x_{i}$ is the $i^{th}$ frame of the signal.
    
    \item \textbf{Filter Banks} are computed using triangular filterbanks that are mel scaled. They are applied in order to capture the energy at critical bands and to approximate the shape of the spectrum. Another important use of this step is to reduce the dimensionality of the extracted features. And as the filterbanks are Mel scaled they tend to be more closely spaced at low frequencies and far apart at higher frequencies. The equations to convert between Hertz (f) and Mel (m) is given in \ref{meltohz}
    
    \begin{equation} \label{meltohz}
    \begin{array}{c}
    m=2595 \log _{10}\left(1+\frac{f}{700}\right) \\
    f=700\left(10^{m / 2595}-1\right)
    \end{array}\end{equation}
    
    And the triangular filterbanks can be obtained using the following equations given in \ref{triangularfilterbank}

    \begin{equation} \label{triangularfilterbank}
    H_{m}(k)=\left\{\begin{array}{cl}
    0 & k<f(m-1) \\
    \frac{k-f(m-1)}{f(m)-f(m-1)} & f(m-1) \leq k \leq f(m) \\
    \frac{f(m+1)-k}{f(m+1)-f(m)} & f(m) \leq k \leq f(m+1) \\
    0 & k>f(m+1)
    \end{array}\right.\end{equation}
    
    where $m$ is the index of mel scaled filters, $k$ is the number of DFT bins and $f()$ is the list of mel scaled frequencies.
    
    \item \textbf{Discrete Cosine Transform (DCT)} is performed over the resultant filterbank coefficients. The spectrum obtained from the above steps have some aspects that are more relevant than others. When DCT is performed and out of the resultant coefficients we keep the lower ones and discard the higher coefficients, it tends to capture the important parts of the signal and discarding the extra noise. Due to the properties in Fourier transform like they are made up of sinusoids which have integer cycles and the Fourier basis function start and end at the same value, it does not do a good job at representing or approximating the real human speech. Whereas the DCT are formed of cosines that have half-integer number of cycles and it does not assume periodic extension it is more suitable for representing the signal.

\end{itemize}

\section{Mel Wavelet Packet Coefficients (MWPC)} \label{MWPC}

\begin{figure}[h]
    \centering
    \includegraphics[scale = 0.6]{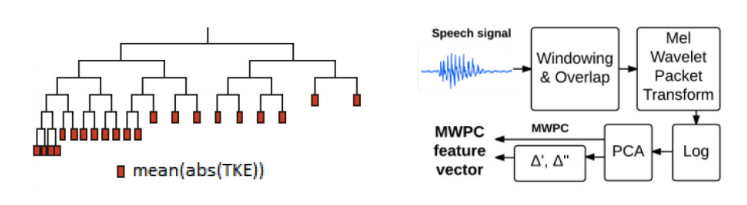}
    \caption{Wavelet packet transform with Teager Keaser Energy nodes on the left and the MWPC implementation on the right.  (\cite{7472724}).}
    \label{fig:WPT_pipeline}
\end{figure}

MWPC gives a detailed time-frequency analysis of the signal that can be used in the countermeasures. It can leverage the multiresolution analysis offered by the wavelet transform to capture the artifacts produced by spoofing techniques. The class of wavelet transform used over here is the wavelet packet transform. A figure given in \ref{fig:WPT_pipeline} for the feature extraction pipeline and brief description explaining the implementation is as follows:

\begin{itemize}
    \item The first few step like \textbf{Pre-emphasis, Framing and windowing} in MWPC are similar to that of MFCC's implementation given in \ref{MFCC}.
    
    \item \textbf{Teager Keiser Energy (TKE)} is calculated for each frame. It is observed that TKE is used for speech quality assessment because of its property to capture more accurate energy that is more useful than the classical sub-band energies, in the field of speech recognition. The computation techniques are more closer to the human auditory system and how it corresponds to noise and meaningful information perceived by the ear. The equation for the calculation of the TKE is given in \ref{TKE} 
    
    \begin{equation} \label{TKE}
    \Psi(s(t))=s(t)^{2}-s(t-1) s(t+1)\end{equation}
    
    where $s(t)$ is the sample of the signal in the frame.
    
    \item \textbf{Wavelet Packet Transform (WPT)} is applied to each of the resultant frame instead of the STFT taken in MFCC feature. WPT is an extended form of Discrete Wavelet Transform (DWT) where the discrete time signal is passed through more filters than DWT. In case of DWT only the lower sub-bands are kept but for WPT both the high and low sub-bands are retained. WPT is preferred over STFT because of its flexibility of choosing / prioritizing the resolution of frequency over that of time. Whereas in case of STFT both the resolution of frequency and time are fixed for the entire analysis. 
    
   \item \textbf{Magnitude and power spectrum} of the features of each frame are obtained. Then \textbf{Mel Scaled Features} are obtained by a simple dot product of the log energies with the mel scaled filterbank matrix.
   
   \item \textbf{Principal Component Analysis (PCA)} is applied to the resultant mel scaled features to derive 12 coefficients that is then fed to the classifier for spoof detection.
    
\end{itemize}

\section{Delta and Double Delta}

The cepstral vectors in the MFCC features are calculated for a small enough time window assuming that the speech is stationary for it. This assumption of stationarity allows the cepstral features to describe the speech only in an instantaneous manner. This way of describing the speech signal would make sense if speech were just concatenated sequence of phonemes, but in reality acoustic signals are more accurately described as a sequence of transitions between phonemes.  

Delta and double delta are one of the methods to capture transitions between the phonemes. They are also called the velocity and acceleration coefficients respectively. And are know for capturing the non-stationarity of speech signal. Delta is the first difference of the signal features, whereas double delta is the second difference.

\begin{equation}
\Delta_{k}=f_{k}-f_{k-1}
\end{equation}

\begin{equation}
\Delta \Delta_{k}=\Delta_{k}-\Delta_{k-1}
\end{equation}

$\Delta$ and $\Delta \Delta$ are the delta and double delta features. $f_k$ is the feature at kth time instant. respectively are found important to achieve
good spoofing detection performance. These delta and double delta features have show to improve the performance of the countermeasure system's \cite{sahidullah2019introduction}. In some cases, the use of only dynamic features is superior to the use of static plus dynamic coefficients \cite{sahidullah2015ssdfeatures}. This is not entirely surprising, since voice conversion and speech synthesis techniques may fail to model the dynamic properties of the speech signals, introducing artefacts that help the discrimination of spoofed signals.

\section{Experiments}

The experiments carried out were focused on three main points:
\begin{itemize}
    \item To compare the performance of wavelet based features as compared to that of MFCC features. 
    \item To explore the concept of deep features in the field of spoof detection.
    \item To test the credibility of handcrafted features on latest spoofing attacks from the ASVspoof 2019 dataset.
\end{itemize}

\subsection{Experimental setup} \label{MFCCMWPCsetup}

The audio used from the ASVspoof 2015 and 2019 dataset was trimmed down to the length of the shortest signal for simplicity. Input to the classifier is the flattened feature vector which is obtained by windowing the previously extracted MFCC or MWPC vector with a context of 7 left and right samples. All the feature extraction techniques used in the experiments that are performed in this chapter are inspired and mixed and matched from these papers (\cite{7472724}) (\cite{Alam2016features}).

\subsubsection{MFCC}

The pre-emphasis filter coefficient ($\alpha$) used was equal to $0.97$. The framing was done with frame length of 25 ms and with a frame shift of 10 ms. For windowing Hamming window function was multipied with each frame. N-point FFT was performed where $N = 512$. 20 filters were used to create the Mel scaled triangular filterbank to extract the frequency bands. The length of the DCT is equal to 20 in order to get a total feature size 60 including the log energies, delta and double delta features). Only dynamic delta and double delta features are used for the task of spoof detection. The above configuration of features seem to work best for the task of spoof detection and are also mentioned in the work \cite{alam2016} and \cite{sahidullah2015ssdfeatures}. The features were extracted from the speech samples which only contained voice speech segments that were extracted using VAD. common mean variance normalization was performed on the feat     

\subsubsection{MWPC}

The pre-emphasis, framing and windowing configurations are same for MWPC. Except for using the classical subband frequencies TKE is computed for each frame to get closer representation to the human auditory system. In MWPC WPT is implemeted using the pywavelets: a python package for wavelet analysis (\cite{Lee2019}). Daubechies class of Wavelet was used as it was the default setting the package. The level of decomposition of the WPT was set to 4 as any level further than that seemed to deteriorate the performance. PCA is applied on the resultant log energies to reduce down the dimensionality to 12 coefficients. Unlike MFCC features extraction process Voice activity detection and common mean variance normalization were not performed as it decreased the performance of the model.  

\subsubsection{Classifier}

DNN is used as a feature generator to extract compact representation and useful information from the long input signal fed to it. These compact representations are called as Bottle-Neck Features (BNF). The DNN used has 5 hidden layers, the first four layers has 1000 neurons with sigmoid activation each. The 5th layer is the bottleneck layer which is a layer with relatively small number of neurons compared to the other layers of the network. To be precise the number of neurons in the BNF layer were 64. The last layer is softmax layer with 2 neurons, one for spoofed other for human class. Once the BNF features are extracted Gaussian Mixture Model (GMM) is used for classification between human and spoofed samples. One GMM was used to model human samples ($M_{h}$) and other was used to model spoofed samples ($M_{s}$) from the training set. Both the GMM had 512 mixture models. Then finally the log-likelihood ratio scores were calculated for the features that were extracted from the development and evaluation set. The formula to calculate the log-likelihood ratio scores is as follows:

 \begin{equation}L L R=\log \left(p\left(O \mid M_{h}\right)\right)-\log \left(p\left(O \mid M_{s}\right)\right)\end{equation}

\renewcommand{\arraystretch}{1.2}
\setlength{\tabcolsep}{3pt}
\begin{table}[t]

\begin{center}

\begin{tabular}{llllll}
\hline
\hline
 System & \multicolumn{2}{c}{$Development$} &  &\multicolumn{2}{c}{$Evaluation$}    \\ 
\cline{2-3} \cline{5-6} 
        & $EER[\%]$  & $tDCF^{min}_{norm}$ &  & $EER[\%]$  &  $tDCF^{min}_{norm}$                                  \\    
\hline
ASSERT & 0 & 0 & & 0.155 & 0.067\\
\hline
STC & 29.73 & 0.71  & & 47.87 & 0.865 \\
\hline
MFCC-BNF & 28.55 & 0.69         &  &  46.21    &  0.826     \\

MWPC-BNF & 25.23  &  0.64  &  &  44.85  &  0.778 \\

\hline
\hline
 
\end{tabular}
\end{center}
\caption {\label{tab:ASVspoof 2019 results} Logical access results of different countermeasures.} 
\end{table}

\renewcommand{\arraystretch}{1.2}
\setlength{\tabcolsep}{3pt}
\begin{table}[t]

\begin{center}

\begin{tabular}{llll}
\hline
\hline
 System & {$Development$} &  & {$Evaluation$}    \\ 
\hline
        &  $EER[\%]$    &  & $EER[\%]$                                    \\    
\hline
DLPCC-BNF & 0.0 &  & 3.33 \\
\hline
STC & 0.008   & & 3.92  \\
\hline
MFCC-BNF & 0.01 &  &  4.35 \\

MWPC-BNF & 0.004  &  &  2.66 \\

\hline
\hline
 
\end{tabular}
\end{center}
\caption {\label{tab:ASVspoof 2015 results} Results for ASVspoof 2015 dataset for different countermeasures.} 
\end{table}

\section{Results and Discussion}

\begin{figure}[tp]
    \centering
    \includegraphics[scale = 0.3]{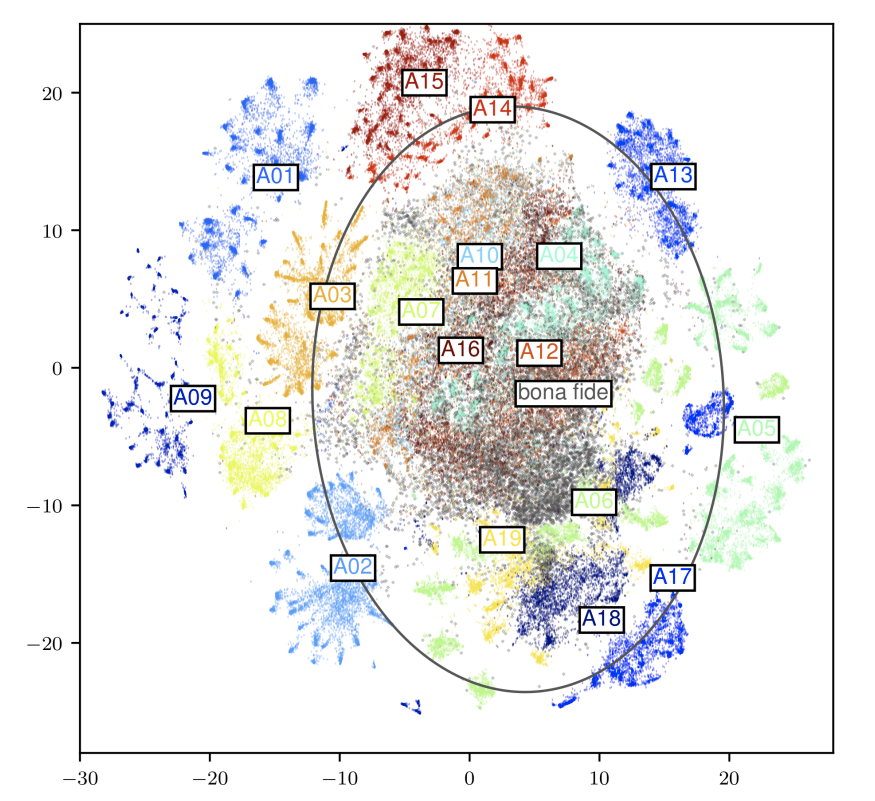}
    \caption{Visualization of bona fide and spoofed speech data of ASVspoof 2019 LA subset. Black circle denotes range of bona fide data within mean $+-3$ standard deviation (\cite{wang2019asvspoof}).}
    \label{fig:visualize_asvspoof2019}
\end{figure}

The EER for MFCC and MWPC deep feature GMM based model for both ASVspoof 2015 and 2019 dataset are given in the table \ref{tab:ASVspoof 2015 results} and \ref{tab:ASVspoof 2019 results} respectively. The first row in both the tables is the state-of-the-art model for the respective dataset. The second row contains the baseline model used for these experiments. Both the baseline model and the SOTA is reffered from the works of (\cite{Alam2016features}) and (\cite{7472724}). The STC model also uses the MWPC features but do not use the deep neural network as their feature generator. And as observed generating features from the DNN does bring some amount of improvement over the baseline model. The MWPC features performed better than MFCC features. This increase in performance of the MWPC is due to the multiresolution analysis capabilities of the WPT features that allow to decompose the speech signal into various frequency bands and offers the flexibility to focus on the artifacts in different frequency regions in the spoofed samples. The results were as expected both the handcrafted models performed brilliantly for ASVspoof 2015 dataset but there was a major decrease in performance for the ASVspoof 2019 dataset. Form the figure given in \ref{fig:visualize_asvspoof2019} it is evident that majority of the samples generated by these spoofing attacks have moved past the line where they can be distinguished from human samples, so it was expected that the traditional handcrafted features like MFCC that were build to model the perception of the human auditory system or features like MWPC for applications like denoising and speech detection in speech would have limited success when put against modern VC and TTS based attacks as these systems have much better capability in modelling the vocal tract shape or the human perceptive model as compared to the traditional GMM based systems. From the above observations and discussion it can be inferred that in order to tackle the spoofing attacks present in the ASVspoof 2019 dataset and the attacks that will be generated with upcoming voice conversion and TTS technologies need a more robust and efficient countermeasure system that can surpass the performance of the traditional handcrafted features. So in the next chapter our experiments will be focused towards building end to end Convolutional neural network based countermeasure systems that not only exploits the knowledge and expertise of signal processing through handcrafted features but also leverage the capabilities of the CNN to learn multiple levels of representation of data.

% Chapter 5

\chapter{CWTnet and WSnet: experimentation on End-to-End spoof detection systems} % Main chapter title

\label{Chapter5} % For referencing the chapter elsewhere, use \ref{Chapter1} 

%----------------------------------------------------------------------------------------

% Define some commands to keep the formatting separated from the content 
% \newcommand{\keyword}[2]{\textbf{#1}}
% \newcommand{\tabhead}[2]{\textbf{#1}}
% \newcommand{\code}[2]{\texttt{#1}}
% \newcommand{\file}[2]{\texttt{\bfseries#1}}
% \newcommand{\option}[2]{\texttt{\itshape#1}}

%----------------------------------------------------------------------------------------

\section{Introduction}

Past and recent research efforts have focused on the selection of appropriate spectral features for spoof detection (\cite{sahidullah2015ssdfeatures}), (\cite{Wu2012DetectingCS}), (\cite{todisco2017cqcc}). One such feature that has worked well for task of spoof detection and forms the subject of our experimental study are Wavelet based features (\cite{7472724}) (\cite{articlespectral}) (\cite{Qian2016deepfeatures}). More recent research has focused on end-to-end deep learning architectures for spoof detection (\cite{8398462, tom2018end}). With this recent direction the classical boundary between a feature extractor and a classifier as separate component is getting increasingly blurred. As of ASVspoof2019 the top 7 submissions towards the LA task and top 6 submissions towards PA task used neural networks. This provides strong encouragement to investigate the use of end-to-end deep learning architectures towards the problem of spoof detection. 

Wavelet based processing has been used across domains such as computer vision (\cite{prakash2018deflecting}) to signal processing (\cite{8404418}) for their multi-resolution analysis capabilities and their application to denoising. (\cite{articlespectral}) studied the application of the Wavelet scattering transform as a front-end for a spoofing countermeasure solution and obtained encouraging results on the ASVSpoof2015 dataset. Furthermore, (\cite{7472724}) used the Wavelet packet transform in their feature ensemble to obtain encouraging results on this data set as well. The Wavelet Scattering transform is known to provide features that are rotation and translation invariant in computer vision (\cite{bruna2012invariant}) while the applicability of the discretized Continuous Wavelet transform to being robust to noise is well known. Since both of these are embeddable in end-to-end deep learning architectures we are motivated to try these as interesting multi-resolution time-frequency front-ends towards the spoofing problem.

%----------------------------------------------------------------------------------------

This chapter introduces the following contributions towards the development of anti-spoof systems:
\begin{itemize}
    \item Initial results on the ASVSpoof2019 dataset on Wavelet based end-to-end architectures. These architectures are based on the Wavelet Scattering Transform and the discretized Continuous Wavelet Transform as feature extractors.
    \item Improvements in performance on the Sincnet architecture baseline~\cite{zeinali2019detecting} when results are fused from the said Wavelet based architectures with the outputs from the Sincnet architecture.
\end{itemize}

The rest of the chapter is organized as follows. Section~\ref{desce2e} presents the details of the end-to-end architectures that are presented in this experimental study. Section~\ref{experiments} presents experimental details and results. Section~\ref{discussion} provides a discussion of the experimental results while Section~\ref{conclusion} concludes the chapter.

\section{Description of end-to-end architectures}\label{desce2e}

\begin{figure}[tp]
    \centering
    \includegraphics[scale = 0.4]{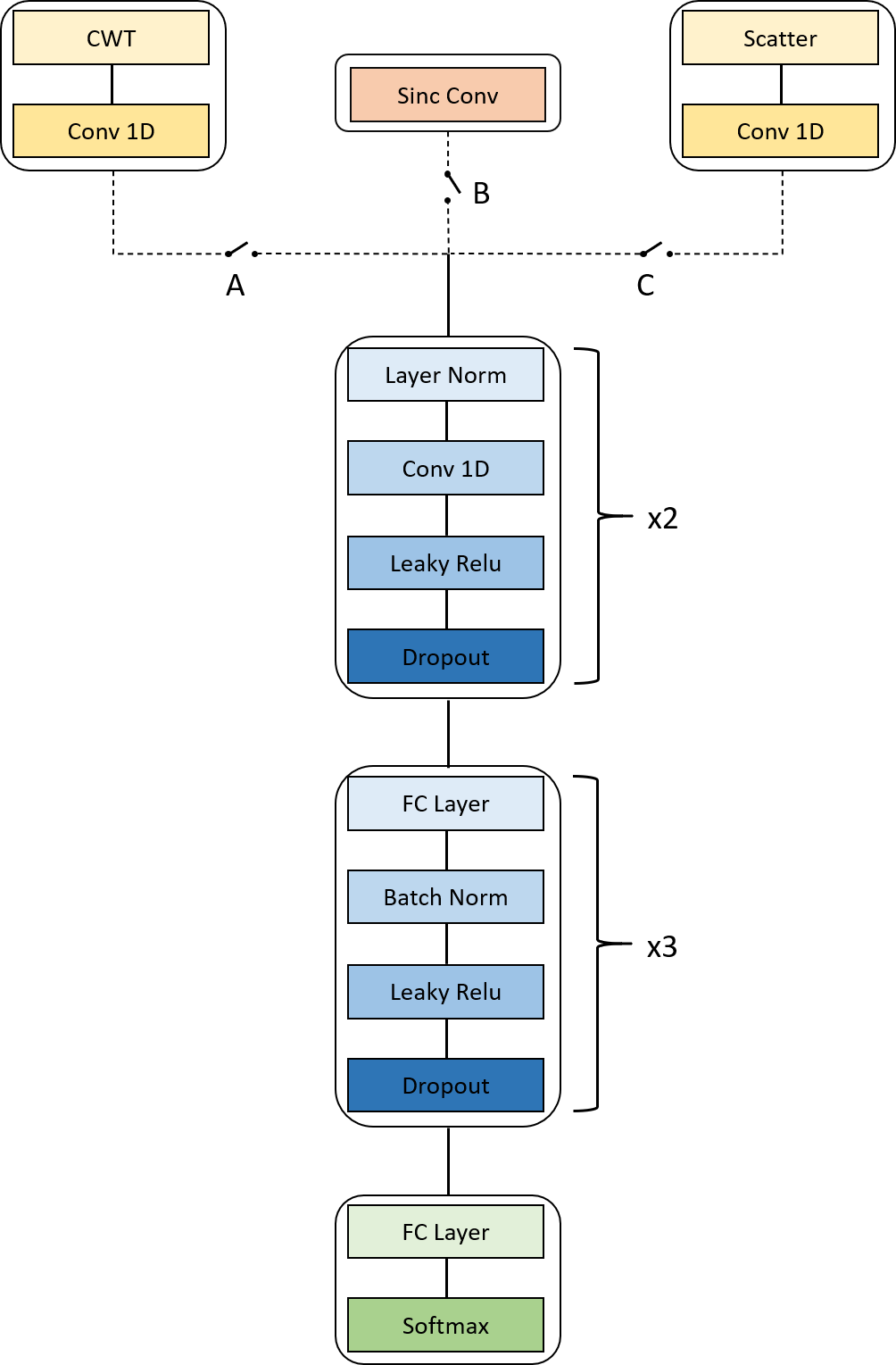}
    \caption{The three architectures :CWTnet, WSnet and Sincnet}
    \label{fig:Fusion_model}
\end{figure}

This section gives a description about the end-to-end architectures that serve as countermeasures for spoof detection. Figure \ref{fig:Fusion_model} gives a brief yet compact representation of three different neural network architecture used in this study. The figure represents alternative paths for each countermeasure. Switch B when connected gives the Sincnet, our baseline end-to-end architecture which is described in Section~\ref{sincnet}. The switch A when connected gives us the discretized Continuous Wavelet Transform Network which we call CWTnet. Switch C when connected gives us the Wavelet Scattering Network which we abbreviate as WSnet. Both of these networks are described in Section~\ref{waveletnets}.      

\renewcommand{\arraystretch}{1.2}
\begin{table}[h]
\caption {\label{tab:table1} Proposed CWTnet architecture. $B$ denotes the batch size, $S$ is the number of scales/channels in the CWT layer. C is the number of channels where $C_1 = (80)$, $C_2 = (80,60)$, $C_3 = 60$. $D$ is the length of the feature map where $D_{inp} = 3200$, $D_1 = 3196$, $D_2 = 1065$, $D_3 = (1065,353)$, $D_4 = (1061,349)$, $D_5 = (353,116)$, $D_6 = 116$, $D_7 = 2048$, $D_8 = 2$  } 
\begin{center}

\begin{tabular}{cccc}
\hline
\hline
Layer Name & Filter &  Output & Blocks    \\ 
\hline 
Input & - & $B$ X 1 X $D_{inp}$ & \multirow{4}{*}{1}\\
CWT & - & $B$ X $S$ X $D_{inp}$ & \\
Conv1d & 5 & $B$ X $C_1$ X $D_1$ & \\
MaxPooling & 3 & $B$ X $C_1$ X $D_2$ & \\
\hline
LayerNorm   & - &  $B$ X $C_2$ X $D_3$ & \multirow{5}{*}{2}  \\
LeakyReLU & -  & $B$ X $C_2$ X $D_3$ &   \\
Dropout & -  & $B$ X $C_2$ X $D_3$ &   \\
Conv1d & 5  & $B$ X $C_3$ X $D_4$  &  \\
MaxPooling & 3 &  $B$ X $C_3$ X $D_5$ & \\
\hline
LayerNorm & -  &  $B$ X $C_3$ X $D_6$  & \multirow{3}{*}{1} \\
LeakyReLU & -  &  $B$ X $C_3$ X $D_6$   & \\
Dropout & -  &  $B$ X $C_3$ X $D_6$   & \\
\hline
Linear  & -  &  $B$ X $D_7$  &   \multirow{4}{*}{3}\\
BatchNorm1d  & -  &  $B$ X $D_7$   &  \\
LeakyReLU & -  &  $B$ X $D_7$  &  \\
Dropout & -  &  $B$ X $D_7$   & \\
\hline
Linear  & -  &  $B$ X $D_8$  &   \multirow{2}{*}{1}\\
Log-Softmax  & -  &  $B$ X $D_8$  &   \\
\hline
\hline
 
\end{tabular}
\end{center}
\end{table}

\renewcommand{\arraystretch}{1.2}
\begin{table}[b!]
\caption {\label{tab:table2} Proposed WSnet architecture. $B$ denotes the batch size. C is the number of channels where $C_1 = (80)$, $C_2 = (80,60)$, $C_3 = 60$. $D$ is the length of the feature map where $D_{inp} = 3200$, $D_1 = 336$, $D_2 = 332$,$D_4 = (332,328)$, $D_4 = (328,324)$, $D_5 = 2048$, $D_6 = 2$ } 
\begin{center}

\begin{tabular}{cccc}
\hline
\hline
Layer Name & Filter &  Output & Blocks    \\ 
\hline 
Input & - & $B$ X 1 X $D_{inp}$ & \multirow{3}{*}{1}\\
WST & - & $B$ X $M$ X $D_1$ & \\
Conv1d & 5 & $B$ X $C_1$ X $D_2$ & \\

\hline
LayerNorm   & - &  $B$ X $C_2$ X $D_3$ & \multirow{4}{*}{2}  \\
LeakyReLU & -  & $B$ X $C_2$ X $D_3$ &   \\
Dropout & -  & $B$ X $C_2$ X $D_3$ &   \\
Conv1d & 5  & $B$ X $C_3$ X $D_4$  &  \\

\hline
LayerNorm & -  &  $B$ X $C_3$ X $D_4$  & \multirow{3}{*}{1} \\
LeakyReLU & -  &  $B$ X $C_3$ X $D_4$   & \\
Dropout & -  &  $B$ X $C_3$ X $D_4$   & \\
\hline
Linear  & -  &  $B$ X $D_5$  &   \multirow{4}{*}{3}\\
BatchNorm1d  & -  &  $B$ X $D_5$   &  \\
LeakyReLU & -  &  $B$ X $D_5$  &  \\
Dropout & -  &  $B$ X $D_5$   & \\
\hline
Linear  & -  &  $B$ X $D_6$  &   \multirow{2}{*}{1}\\
Log-Softmax  & -  &  $B$ X $D_6$  &   \\
\hline
\hline
 
\end{tabular}
\end{center}
\end{table}
\subsection{Sincnet}\label{sincnet}

Sincnet~\cite{ravanelli2018speaker,zeinali2019detecting}
is an end-to-end convolutional neural network architecture for raw audio processing. This novel architecture has proved its credibility in all sorts of raw audio processing tasks ranging from speech recognition to spoof detection. The core of the architecture is its first layer that is based on a series parameterized sinc functions. It implements band-pass filters and is responsible for learning meaningful filters. This prior knowledge drastically reduces the number of trainable parameters relative to normal convolutional layers.

\subsubsection{Model Description}

\begin{itemize}
\item First layer performs the sinc based convolutions. It has 80 filters of length 251 samples. These filters are initialized using Mel frequency filter banks. And the only learnable parameters are the low and high cut off frequencies.  
\item The next two layers are standard convolution layers with 60 filters of length 5.
\item Then comes the 3 fully connected layers of length 2048.
\item The convolutional layers are layer normalized and the FC layers are batch normalized. 
\item Throughout the model Leaky Relu is used as the activation except the last layer where log-softmax is used.
\item Negative log-likelihood loss criterion was used to improve the model.

\end{itemize}

\subsection{CWTnet and WSnet}\label{waveletnets}

CWTnet and WSnet  have exactly the same architecture as Sincnet except the first layer. The sinc convolution layer has been replaced with continuous wavelet transform layer in case of CWTnet and scattering transform layer in case of WSnet. Both the models take raw audio as input. 

\subsubsection{CWTNet} \label{CWTnet}
For CWTnet the first layer returns the discrete continuous wavelet transform as the output to the next layer. Table~\ref{tab:table1} provides the architectural details of the CWTNet. The table has 4 columns namely the layer name, the second column describes the filter size (if applicable), the third column described the output dimension. For ease of representation a group of layers is categorized into a block which is evidenced by the last column of the table (Blocks) which denotes the number of successive blocks in the architecture. In the caption for the table $C_2 = (80, 60)$ means that the parameter $C_{2}=80$ in the first block that it occurred and $C_{2}=60$ in the next successive block. With reference to the CWT layer (in Block 1) the output of the layer is three dimensional and is of the form $[B, S, D_{inp}]$. Here $B$ is the batch size , $S$ is the number of scales over which the CWT is calculated and $D_{inp}$ is the length of the CWT coefficients at every scale. In this work we used the second Derivateive of Gaussian (DoG) or Mexican hat wavelet as the mother Wavelet function, mentioned in the equation as follows:

\begin{equation} \label{DOGeq}
\psi(t)=\frac{2}{\pi^{1 / 4} \sqrt{3 \sigma}}\left(\frac{t^{2}}{\sigma^{2}}-1\right) e^{-\frac{t^{2}}{\sigma^{2}}}\end{equation}

For reference the discretized CWT of a signal $x[n]$ be written as:
\begin{equation} \label{CWTeq}
    \sum_{n}\frac{1}{\sqrt{s_{j}}}\psi(\frac{n-m}{s_j})x[n]
\end{equation}
Here $\Psi$ is a Wavelet and  $m$ is used to denoted the denote the discretized translation parameter. 
The convolution mentioned is carried out over the support of the signal $x[n]$. As mentioned in~\cite{Torrence1998APG} $\psi$  is scaled by  
\begin{equation} \label{eq2}
    \ s_j = s_0 2^{j \delta j} 
\end{equation}
\begin{equation}\label{eq3}
    \ J = \delta j^{-1} log_2 (N n / s_0) 
\end{equation}
Here $s_0$ is the smallest resolvable scale, $J$ determines the largest scale and $n$ is the time step of the signal of length $N = D_{inp}$. The $s_0$ should be chosen so that the equivalent Fourier period is approximately $2 n$. $\delta j$ is the scale distribution parameter and it depends on the width in spectral-space of the wavelet function (\cite{Torrence1998APG}). In this work the value of $\delta j$ and $n$ is chosen to be 0.125 and 0.1 respectively. The optimum scale distribution is then used to initialize the wavelet filter-bank. This filter-bank is implemented using a collection of 1d convolution filters which are trainable in nature. These filters takes batch of signals and convolves each signal with all elements in the filter-bank. After convolving the entire filter bank, the method returns the final output of the mentioned above.

\subsubsection{WSNet} 
In case of WSnet the first layer takes three inputs (a) the signal length (b) $N_{k}$ the number of wavelets per level and (c) the $Q_k$ number of wavelets per octave. The reader is referred to~\cite{Anden_2014} for details of the Wavelet Scattering Transform. It is to be noted that the parameters in this layer are not trainable. The Kymatio Python library~\cite{andreux2020kymatio} was used to implement the Wavelet Scattering transform for the WSNet feature extraction layer. Our implementation used 2 levels of Wavelet Scattering with $N_{1}=12$ and $N_{2}=1$ and $Q_{1}=Q_{2}=8$ across the two levels. The choice of the Wavelet function was set to be the at the default setting in this library.

In summary the scattering transform works in an iterative fashion where $N_k$, $\phi_k$ are the number of filters and scaling function at $k^{th}$ level of decomposition. At level $k$ the Wavelet filters are notated as:
\begin{equation}
\psi_{k_j}[n] ; j = 1,2,3,...,N_k
\end{equation}
More specifically, the dilated Wavelet functions are of the form:
\begin{equation}
    \psi_{k_j}[n]  =  \frac{1}{s}  \psi_k [\frac{nT}{s}] , s = 2^{-j/Q_k}
\end{equation}
Here, $s$ is the scaling factor; $T$ is the sampling period; $n$ is the sample index; $Q_k$ is the number of wavelets in an octave at $k^{th}$ level of decomposition and $x[n]$ denotes the input signal.At each level of the decomposition one obtains \textit{scalogram} coefficients by computed by taking the absolute values of the outputs of the $N_{k}$ filters and is given by the set of signals ${p_j[n]}$ where:
\begin{equation}
    p_j[n] = |(\psi_j * x)[n]|
\end{equation}
Here  $| . |$ denotes the modulus operator. The \textit{scattering} coefficients at each level are obtained by framing $p_j[n]$ using rectangular windows (scaling functions) $\phi_{k}$ of length $M$ and taking the average value within these frames to obtain the $j^{th}$ scattering coefficient corresponding to the signal $x[n]$ at the $k^{th}$ level of decomposition, i.e.:
\begin{equation}
    S_k^{(j)}[m] = ( \phi_k * p_j ) [mM] ; \text{for } k > 1
\end{equation}
where, $S_k^{(j)}[m]$ denotes the scattering coefficient corresponding to the $j^{th}$ bandpass filter $\psi_{j}[n]$ at the $k^{th}$ level of decomposition and $m$ denotes the frame index. The $0^{th}$ order scattering coefficients are obtained by convolving the raw audio input with the scaling function at this level .

We remove the zeroth order scattering coefficients as it did not make any significant difference in the results. And to improve the discriminability we take the logarithm of the scattering coefficients (after adding a small constant to make sure nothing blows up when scattering coefficients are close to zero). This is known as the log-scattering transform. The architectures of WSnet is given in Table \ref{tab:table2}. The table is formatted in a similar manner to Table\ref{tab:table1}. The output of the WST layer is three dimensional and is of the form $[B, M, D_{1}]$. Here $B$ is the batch size , $M$ is the number scattering coefficients and $D_{1}$ is the sub-sampled signal length.

\section{Experiments}\label{experiments}

In this section we report our results for development and evaluation set from both Logical access part of the ASVspoof2019 dataset.

\subsection{Dataset and Performance metric}

All our models are trained and evaluated strictly on the ASVspoof2019 Dataset. The dataset encompasses two partitions for the analysis of Logical Access and Physical Access. Both LA and PA are themselves divided into three partitions, namely training, development and evaluation. Each of these sets are disjoint in terms of speakers. Our experiments present results only on the logical access portion of the dataset. These spoofed speech samples are synthesized using 6 different state-of-the-art text-to-speech or voice-conversion models. 

The primary mode to evaluate the countermeasures is the min-tDCF metric~\cite{kinnunen2018t} which takes into consideration the ASV system's performance along with the countermeasure (Sincnet, CWnet and WSNet) system's performance and the equal-error rate is used as the secondary mode of evaluation.

\begin{figure}[h]
  \centering
   \caption{Histogram and min-tdcf curve for CWTnet, WSnet and fusion-1 model. Top row is for CWTnet, second row is for WSnet and last row is for fusion-1 model}
  \begin{minipage}[b]{0.3\textwidth}
    \includegraphics[width=\textwidth]{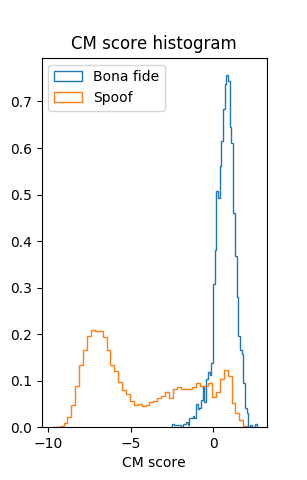}
  \end{minipage}
  \hfill
  \begin{minipage}[b]{0.6\textwidth}
    \includegraphics[width=\textwidth]{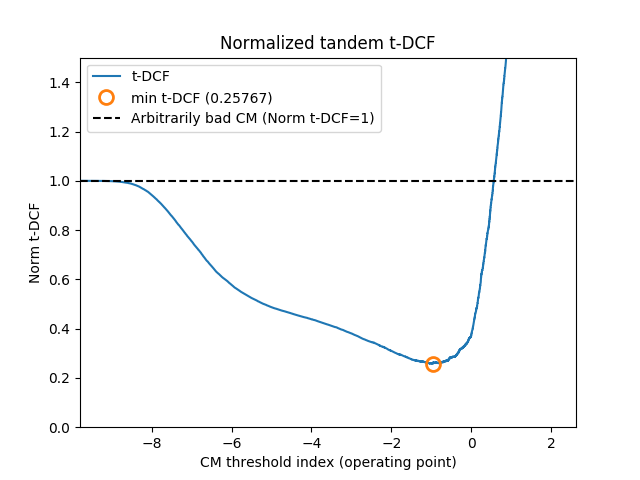}
  \end{minipage}
  \begin{minipage}[b]{0.3\textwidth}
    \includegraphics[width=\textwidth]{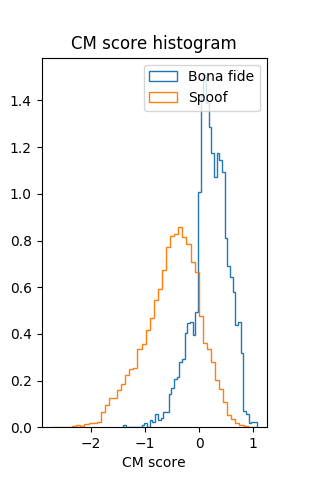}
  \end{minipage}
  \hfill
  \begin{minipage}[b]{0.6\textwidth}
    \includegraphics[width=\textwidth]{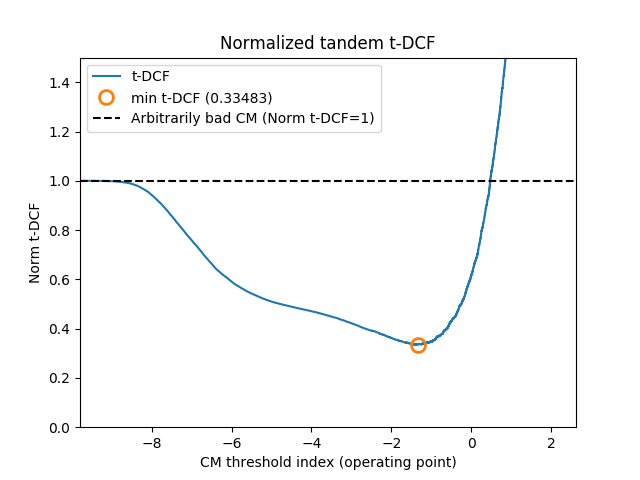}
  \end{minipage}
  \begin{minipage}[b]{0.3\textwidth}
    \includegraphics[width=\textwidth]{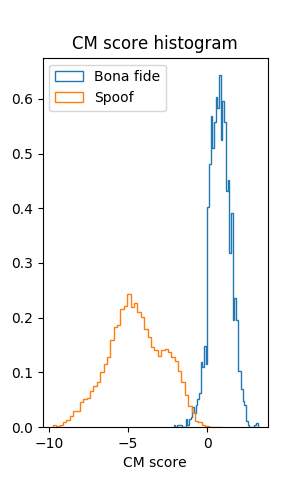}
  \end{minipage}
  \hfill
  \begin{minipage}[b]{0.6\textwidth}
    \includegraphics[width=\textwidth]{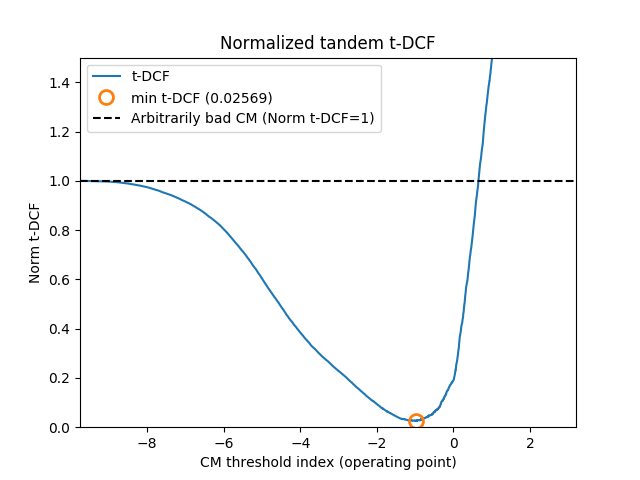}
  \end{minipage}
\label{fig:histogram and mintdcf}
\end{figure}

\subsection{Methodology used for training the countermeasures} \label{CWTtraining}
The learning of countermeasure systems is highly dependent on how we feed the data to it. All the countermeasure systems are fed in a similar fashion. Out of the whole training set 256 utterances are picked at random and from each of those utterances random chunks of 200ms is cropped to form a mini-batch which is fed to the model at every step~\cite{zeinali2019detecting}. Every mini-batch has equal number of spoofed and bonafide samples where every spoofed sample has a bonafide sample of the same speaker. This way we make sure that the model is only learning to distinguish between spoofed and bonafide samples and is not biased towards the majority class. It also ensures that the models learning is not getting confused by speaker characteristics of various speakers. We used RMSprop optimizer to change our weights. The model was seen to train well with the use of the Log-softmax at the output layer with negative log-likelihood loss function. 

\subsection{Methodology for evaluation} \label{E2Eevaluation}

The model is trained, validated and tested on the ASVspoof 2019 train,  development and evaluation set respectively. The validation and evaluation is done at an utterance level, that is, score is calculated for every frame and are averaged for the whole utterance. The final score value along with all the other scores values for the utterances in the dataset are used for calculating the EER[\%] and tDCF scores. The model is trained for 50 epochs. The EER[\%] score is calculated after ever epoch for the development set, weights are saved if EER[\%] score improves. The best model weights are then used for evaluation and finally the EER[\%] and tDCF values are reported in this document. 

\subsection{Results}

The result of our countermeasures for LA are reported in Table \ref{tab:table3}. For reference, the first row reports the state-of-the-art results on this dataset obtained by the JHU ASSERT system~\cite{Lai2019ASSERTAW}. The second row reports results using the Sincnet which serves as our baseline. This baseline was repeated by us and is consistent with the results reported by~\cite{zeinali2019detecting}. The results of CWTnet and WSnet are given in the third and fourth row respectively. And the last three rows give the results of fusing our Sincnet baseline model outputs with the CWTnet and the WSNet in various combinations using weighted averaging. Details of the fusion combinations are as below:

\begin{itemize}
    \item \textbf{Fusion - 1} : Fusion of CWTnet, Sincnet and WSnet. The weighted average is calculated by giving 75\% weight to Sincnet output and 25\% weight is equally divided between the other two models.
    \item \textbf{Fusion - 2} : Fusion of CWTnet and Sincnet with 75\% weight given to Sincnet and 25\% weight given to CWTnet.
    \item \textbf{Fusion - 3} : Fusion of WSnet and Sincnet with 75\% weight given to Sincnet and 25\% weight given to WSnet.
\end{itemize}

Figure \ref{fig:histogram and mintdcf} contains the histogram and the min-tdcf curve for CWTnet, WSnet and fusion-1 model. The histogram gives us an idea about how confident the model was in distinguishing spoof and bonafide samples by plotting the density of CM score values. Form the histogram of all the models we can observe that the models are quite confident about most of the bonafide samples and have high number of CM scores between the value of 0 and 1. In case of spoof samples the CWTnet seems to be confident about classifying some of the spoofing attacks and hence has moderate amount of CM scores far apart for the CM scores of bonafide samples. The WSnet seems to barely distinguish between the spoof and bonafide samples which is evident from the high number of CM scores of both spoof and bonafide samples that are not the far apart. The fusion model seems to work the best and has well defined peaks for both spoof and bonafide samples that are far apart.     

\renewcommand{\arraystretch}{1.2}
\setlength{\tabcolsep}{3pt}
\begin{table}[t]
\caption {\label{tab:table3} Logical access results of different countermeasures.} 
\begin{center}

\begin{tabular}{llllll}
\hline
\hline
 System & \multicolumn{2}{c}{$Development$} &  &\multicolumn{2}{c}{$Evaluation$}    \\ 
\cline{2-3} \cline{5-6} 
        &  $EER[\%]$  & $tDCF^{min}_{norm}$ &  & $EER[\%]$  &  $tDCF^{min}_{norm}$                                  \\    
\hline
ASSERT & 0 & 0 & & 0.155 & 0.067\\
\hline
Sincnet & 1.45 & 0.04  & & 20.11 & 0.358 \\
\hline
CWTnet & 12.56 & 0.25         &  &  25.42    &  0.508     \\
% CWTnet-F & 1.47 & 0.012         &  &  19.36    &  0.320   \\
WSnet & 18.74  &  0.33  &  &  39.13  &  0.646 \\
% WSnet-F & 1.60 & 0.013 & & 20.13 &  0.355 \\
\hline
\textbf{Fusion-1} & \textbf{0.82} & \textbf{0.027} &  &  \textbf{16.67} & \textbf{0.269} \\ 
\textbf{Fusion-2} & \textbf{1.13} & \textbf{0.038} &  &  \textbf{18.45} & \textbf{0.322} \\ 
\textbf{Fusion-3} & \textbf{1.373} & \textbf{0.034} &  &  \textbf{19.22} & \textbf{0.329} \\ 
\hline
\hline
 
\end{tabular}
\end{center}
\end{table}

\section{Discussion}~\label{discussion}

From the results we can infer that our individual models CWTnet and WSnet are not up to the mark at the task of spoof detection. While we use the Sincnet model as our baseline, it is apparent that its performance is lacking when compared to the state of the art. Signifcant improvements to this baseline are reported with the use of fusion of results from our wavelet based models with the results of the Sincnet. Fusion-1 which involves a combination of the CWTnet, WSNet, and the Sincnet gave the best results. While the Fusion-1 model gives comparable results to the state-of-the-art on development set it fails to generalize on the evaluation set. As expected Fusion-2 (CWTNet + Sincnet) and Fusion-3 (WSNet + Sincnet) models do provide improvements over the Sincnet baseline in manner that is consistent with the individual models' (CWTNet and SincNet) performance.

These results are encouraging and point to the fact that spectral diversity at the input feature level is an asset. The results from Fusion-1 seem to indicate that the Wavelet based models are able to capture what the SincNet analysis seems to miss. The results are also able to tell us the amount of research work remaining to adapt a powerful formalism such as Wavelets into end-to-end architectures for spoofing. We do recognize that optimal fusion weights could be determined by tuning on the development set. This will be presented in future work. Furthermore, we plan to extend the use of 
attention filtering networks~\cite{Lai2019ASSERTAW} to help enhance the features in certain frequency regions for better results.

\section{Conclusions}\label{conclusion}

This chapter presented an initial experimental investigation of Wavelet based feature extractors such as the Wavelet Scattering transform and the discrete Continuous Wavelet Transform embedded into end-to-end architectures. While the performance of the standalone models was not encouraging it was seen that fusion with results from the Sincnet model provided significant improvements over the Sincnet baseline. It is apparent that these models are able to capture something significant that the Sincnet models seem to miss. 
% Future work would involve the development of novel spectral front-ends in end-to-end neural network based architectures. 

% Chapter 6

\chapter{Towards Learn-able Parametric Filters for Spoof Detection} % Main chapter title

\label{Chapter6} % For referencing the chapter elsewhere, use \ref{Chapter1} 

%----------------------------------------------------------------------------------------

% Define some commands to keep the formatting separated from the content 
% \newcommand{\keyword}[2]{\textbf{#1}}
% \newcommand{\tabhead}[2]{\textbf{#1}}
% \newcommand{\code}[2]{\texttt{#1}}
% \newcommand{\file}[2]{\texttt{\bfseries#1}}
% \newcommand{\option}[2]{\texttt{\itshape#1}}

%----------------------------------------------------------------------------------------

\section{Introduction}

From our experiments in the previous chapters it is observed that spectral decomposition using handcrafted features when combined with deep neural networks seem to improve the performance as compared to the traditional spoof detection models. Even though the models CWTnet and WSnet given in section \ref{waveletnets} with minor fine-tuning of hyper-parameters performed better than MFCC-BNF and MWPC-BNF in section \ref{MFCCMWPCsetup} their performance is far from the SOTA for this task as seen in the table \ref{tab:table3}. Having observed that wavelet based features seem to have limited success in the field of spoof detection it would not be preferable to put more effort in trying out different architectures on top of the CWT or WS layer. So this chapter only mildly focuses on improving the performance through feature selection and mostly focuses on  using the proposed method given by \cite{NIPS2018_7711} to get idea about the frequencies that supposedly contain artifacts and are responsible for spoof detection. 

In section \ref{CWTnet} the scale distribution and the number of scales used to extract features in CWTnet was decided on the basis of the width of the spectral-space of the wavelet function and the length of the signal. Obviously the final scale distribution and number of scales obtained are far from optimal for the task at hand. One way effective spectral decomposition is to fine tune the hyper-parameters like these using wrapper methods where one would have to train the model multiple times using different feature sets and then compare the resulting models via their cross validation accuracy. Another way would be to have some knowledge about the relevant frequency ranges that are present in the input signal and would help in discriminating between human and spoofed samples and using that information one could manually experiment by setting appropriate number of scales and distribution. But both these methods would be very computationally and time expensive. Different hyper-parameters for spectral decomposition are used for different time series based problems and as discussed above optimizing these hyper-parameters is not an easy task. The idea given in \cite{NIPS2018_7711} precisely provides a solution for this problem. It proposes a method called wavelet deconvolution to efficiently optimize the hyper-parameters that is the scale parameter based on discrete form of continuous wavelet transform in a neural network framework using back-propagation. 

The rest of the chapter is organised as follows. Section \ref{WD} describes about the wavelet deconvolution method for learning the scale parameter and relevant background. Section \ref{WDExp} cover all the experiments. Section \ref{WDRes} talks about the results achieved by this method and discussions related to it.

\section{Wavelet Deconvolution (WD)} \label{WD}

The Mexican hat wavelet is used as the mother wavelet function, its equation and the CWT of the signal is given in equation \ref{DOGeq} and \ref{CWTeq} respectively. The Mexican hat wavelet is chosen as the mother wavelet function because of its compact support which helps in local analysis of the signal and the integral or sum can be easily performed while calculating the wavelet transform. But in order to maintain consistency in this chapter its worth writing the scaled wavelet function and CWT equation in the following form:

\begin{equation}\psi_{s, b}(t)=\frac{1}{\sqrt{s}} \psi\left(\frac{t-b}{s}\right)\end{equation}

\begin{equation}W_{x}(s, b)= \sum_{t} \frac{1}{\sqrt{s}} \psi\left(\frac{t-b}{s}\right) x(t)\end{equation}

Here $\psi_{s,b}$ is the wavelet function scaled at $s$ and translated at $b$. $W_{x}(s,b)$ is the discrete continuous wavelet transform. As mentioned in the paper, choosing the values for the scale parameter can be assisted by converting scales to frequencies. This can be done using the center frequency of the mother wavelet. The equation for converting the scales to frequencies is given by :

\begin{equation} \label{freqtoscale}
F_{s}=\frac{F_{c}}{s}\end{equation}

Here $F_{c}$ is the center frequency of the mother wavelet, $F_{s}$ is the frequency at scale $s$. But this would not be of much help as the relevant frequency content which is responsible for distinguishing the spoofed samples from the human voice are unknown. \cite{NIPS2018_7711} provides a convolutional neural network based setting the replaces the preprocessing step that preforms the wavelet transform with a layer called the wavelet deconvolution(WD) layer. This layer calculates the discrete continuous wavelet transform similar to the CWTnet but also optimizes the scale parameter using back-propagation. WD layer calculates the transform on the input signal during the forward pass, gives the resultant features to the further layers of the network. It computes the gradients of the loss function with respect to the scale parameters during back-propagation. A few reason to modify the normal CWT layer to a WD layer is as follows:

\begin{itemize}
    \item It gives an insight about the relevant frequency content that is responsible for distinguishing between spoof and human samples.
    \item From the learned scale parameters the kernel size can be implicitly adapted providing multi-resolutional analysis at selected frequency regions.
    \item More generalized features are extracted as compared to the features extracted by CWTnet as only the most important and relevant frequency regions are focused upon. 
    \item It resembles a feature selection approach due to which improvement is observed in training speed and EER as compared to CWTnet. And it is better than the expensive and time consuming cross validation based feature selection methods.
    
\end{itemize}

Here is a brief description about the calculation of gradients which is done while optimizing the scale parameters. In this case the signal in ASVspoof 2019 database has only single channel but it is applicable to multi-channel cases as well by applying transform to every channel. First, forward pass through the WD layer is done on the signal as represented below:

\begin{equation}z_{i}=x * \psi_{s_{i}} \forall i=1 \ldots M\end{equation}

This equation of convolution can also be written in the form of summation:

\begin{equation}\begin{array}{c}
z_{i j}=\sum_{k=1}^{K} \psi_{s_{i}, k} x_{j+k} \\
\text { for } i=1 \ldots M \text { and } j=1 \ldots N
\end{array}\end{equation}

The signal $x \in R^{N}$ where $N$ is the number of samples. Scale is denoted by $s \in R^{M}$ where $s>0$. After convolving the wavelet function $\psi_{s_{i}}$ that is scaled at  $s_{i}$ we obtain $z_{ij} \in R^{MXN}$. The $\psi_{s_{i}}$ is parametrized over $K$ points.

\begin{equation}\begin{array}{l}
\psi_{s_{i}, t}=\frac{2}{\pi^{\frac{1}{4}} \sqrt{3 s_{i}}}\left(\frac{t^{2}}{s_{i}^{2}}-1\right) e^{-\frac{t^{2}}{s_{i}^{2}}} \\
t \in\left\{-\frac{K-1}{2}, \ldots 0, \ldots \frac{K-1}{2}\right\}
\end{array}\end{equation}

The scale parameter is differentiable with respect to the error function hence it can be optimized using back-propagation while minimizing the error. Here is a step-by-step calculation for the whole process:

\begin{equation} \label{derivativewrts}
\frac{\delta E}{\delta s_{i}}=\sum_{k=1}^{K} \frac{\delta E}{\delta \psi_{s_{i}, k}} \frac{\delta \psi_{s_{i}, k}}{\delta s_{i}}\end{equation}

Here $E$ is the differentiable loss function that needs to be minimized. $\frac{\delta E}{\delta s_{i}}$ is the partial derivative of $E$ w.r.t. each scale parameter $s_{i}$. The equation \ref{derivativewrts} can be broken down into two parts one related to the gradients w.r.t. to the filter $\psi_{s_{i},k}$ and other for the gradient of $\psi_{s_{i},k}$ w.r.t. the scale $s_{i}$.

\begin{equation} \label{gradientwrtpsi}
\frac{\delta E}{\delta \psi_{s_{i}, k}}=\sum_{j=1}^{N} \frac{\delta E}{\delta z_{i j}} \frac{\delta z_{i j}}{\delta \psi_{s_{i}, k}}=\sum_{j=1}^{N} \frac{\delta E}{\delta z_{i j}} x_{j+k}\end{equation}

The above equation \ref{gradientwrtpsi} is for the gradient w.r.t. the filter which is written using $\frac{\delta E}{\delta z_{i j}}$ which is the gradient w.r.t. to the output of the WD layer.

\begin{equation} \label{splitnpd}
A=\frac{2}{\pi^{\frac{1}{4}} \sqrt{3 s_{i}}}, \frac{\delta A}{\delta s_{i}}=-\frac{3}{\pi^{\frac{1}{4}}}\left(3 s_{i}\right)^{-\frac{3}{2}} M=\left(\frac{t_{k}^{2}}{s_{i}^{2}}-1\right), \frac{\delta M}{\delta s_{i}}=-\frac{2 t_{k}^{2}}{s_{i}^{3}} G=e^{-\frac{t_{k}^{2}}{s_{i}^{2}}}, \frac{\delta G}{\delta s_{i}}=\frac{2 t_{k}^{2}}{s_{i}^{3}} e^{-\frac{t_{k}^{2}}{s^{2}}}\end{equation}

\begin{equation} \label{pdassemble}
\frac{\delta \psi_{s_{i}, k}}{\delta s_{i}}=A\left(M \frac{\delta G}{\delta s_{i}}+G \frac{\delta M}{\delta s_{i}}\right)+M G \frac{\delta A}{\delta s_{i}}\end{equation}

The above two equations is related to the gradient of the filter w.r.t. the scale parameter. In equation \ref{splitnpd} the filter function is broken down into 3 parts $A,M,G$ for convenience and their partial derivative is calculated. In equation \ref{pdassemble} $A,M,G$ and their partial derivatives are written in a proper manner to give the gradient.

\begin{equation} \label{derivativewrtsfinal}
\frac{\delta E}{\delta s_{i}}=\sum_{k=1}^{K}\left[\left(\frac{4 t_{k}^{4}}{s_{i}^{4}}-\frac{9 t_{k}^{2}}{s_{i}^{2}}+1\right) \frac{e^{-\frac{t_{k}^{2}}{s_{i}^{2}}}}{\pi^{\frac{1}{4}} \sqrt{3 s_{i}^{3}}}\right] \sum_{j=1}^{N} \frac{\delta E}{\delta z_{i j}} x_{j+k}\end{equation}

The Equation \ref{derivativewrtsfinal} finally gives the gradients of the loss function w.r.t. to the scale parameter that are used to update the scales using the gradient descent step given in equation below where $\gamma$ is the learning rate and and $s_{i} > 0$ as this ensures the validity of the wavelet function.

\begin{equation}s_{i}^{\prime}=s_{i}-\gamma \frac{\delta E}{\delta s_{i}}\end{equation}

\section{Experiments} \label{WDExp}

This section describes the implementation of the neural network architecture with wavelet deconvolution layer as the first layer. Experiments were carried out with 8 and 20 scales in the WD layer. The scale parameters $s_{1} \ldots s_{M}$ where $M$ is 8 or 20, are initialized using didactic values (1,2,4,8,16,32,64,128 and so on) as the relevant frequency content for the task is unknown.  This is done by creating non-overlapping frequency bins where the number of bins are equal to the number of scales. For fair comparison the neural network architecture that was fed with the WD layer output and helped optimize the scale parameter for extracting relevant features for spoof detection was similar to that used in CWTnet given in \ref{CWTnet}. The loss function used is the categorical cross entropy and the optimization algorithm used is Mean Squared Error(MSE).The model was trained, validated and tested using the provided partitions of the ASVspoof2019 database. The data feeding and training mechanism is also kept the same as mentioned in section \ref{CWTtraining}.           

% \renewcommand{\arraystretch}{1.2}
% \setlength{\tabcolsep}{3pt}
% \begin{table}[t]
% \caption {\label{tab:WDresult} Logical access results of different countermeasures.} 
% \begin{center}

% \begin{tabular}{llllll}
% \hline
% \hline
%  System & \multicolumn{2}{c}{$Development$} &  &\multicolumn{2}{c}{$Evaluation$}    \\ 
% \cline{2-3} \cline{5-6} 
%         &  EER[\%]  & tDCF^{min}_{norm} &  & EER[\%]  &  tDCF^{min}_{norm}                                  \\    
% \hline
% ASSERT & 0 & 0 & & 0.155 & 0.067\\
% \hline
% CWTnet & 12.56 & 0.25         &  &  25.42    &  0.508     \\
% % CWTnet-F & 1.47 & 0.012         &  &  19.36    &  0.320   \\
% WSnet & 18.74  &  0.33  &  &  39.13  &  0.646 \\
% % WSnet-F & 1.60 & 0.013 & & 20.13 &  0.355 \\
% \hline
% WavDeconv-8 & 8.36 & 0.20 & & 18.80 & 0.29 \\
% WaveDeconv-20 & 11.34 & 0.26 & & 22.57 & 0.33 \\
% \hline
% \hline
 
% \end{tabular}
% \end{center}
% \end{table}

\renewcommand{\arraystretch}{1.2}
\setlength{\tabcolsep}{3pt}
\begin{table}[t]
\caption {\label{tab:WDresult} Logical access results of different countermeasures.} 
\begin{center}
\begin{tabular}{llllll}
\hline
\hline
 System & \multicolumn{2}{c}{$Development$} &  &\multicolumn{2}{c}{$Evaluation$}    \\ 
\cline{2-3} \cline{5-6} 
        &  $EER[\%]$  & $tDCF^{min}_{norm}$ &  & $EER[\%]$  &  $tDCF^{min}_{norm}$\\    
\hline
ASSERT & 0 & 0 & & 0.155 & 0.067\\
\hline
Sincnet & 1.45 & 0.04  & & 20.11 & 0.358 \\
\hline
CWTnet & 12.56 & 0.25         &  &  25.42    &  0.508     \\
% CWTnet-F & 1.47 & 0.012         &  &  19.36    &  0.320   \\
WSnet & 20.14  &  0.36  &  &  39.13  &  0.646 \\
% WSnet-F & 1.60 & 0.013 & & 20.13 &  0.355 \\
\textbf{WavDeconv-8} & \textbf{8.36} & \textbf{0.20} & & \textbf{18.80} & \textbf{0.291} \\
WavDeconv-20 & 11.34 & 0.24 & & 22.57 & 0.334  \\
\hline
\textbf{Fusion-1} & \textbf{0.9} & \textbf{0.003} &  &  \textbf{16.67} & \textbf{0.269} \\ 
Fusion-2 & 1.334 & 0.012 &  &  18.45 & 0.322 \\ 
Fusion-3 & 1.373 & 0.014 &  &  19.22 & 0.329 \\ 
\hline
\hline
 
\end{tabular}
\end{center}
\end{table}

\section{Result and Discussion} \label{WDRes}

% \begin{figure}[h]
%   \centering
%   \caption{The visualizations of the 8 and 20 wavelet scales and the frequency response of the created filterbank using the wavelet filters. Top row is for WavDeconv-8 and bottom row is for WavDeconv-20}
%   \begin{minipage}[b]{0.3\textwidth}
%     \includegraphics[width=\textwidth]{Figures/scale_8.png}
%   \end{minipage}
%   \hfill
%   \begin{minipage}[b]{0.5\textwidth}
%     \includegraphics[width=\textwidth]{Figures/deconv_8_freq_response.png}
%   \end{minipage}
%   \begin{minipage}[b]{0.3\textwidth}
%     \includegraphics[width=\textwidth]{Figures/scale_20.png}
%   \end{minipage}
%   \hfill
%   \begin{minipage}[b]{0.5\textwidth}
%     \includegraphics[width=\textwidth]{Figures/deconv_20_freq_response.png}
%   \end{minipage}
% \label{fig:Wavdeconv scale and frequency response}
% \end{figure}

The result of the Wavelet deconvolution layer with 8 and 20 scales are given in the table \ref{tab:WDresult} and the hisogram and min-tdcf curve for the models are given in \ref{fig:Wavdeconv histogram and mintdcf}. The first row of the table states the SOTA in ASVspoof2019 that is ASSERT countermeasure. The second and third row states the baseline model(CWTnet and WSnet) upon which our proposed model improves. And the last two rows states the result for wavelet deconvolution models(WavDeconv-8 and WavDeconv-20). It can be observed from the results that the WD layer even though is not close to the SOTA but still have brought major improvements over CWTnet and WSnet models. Another point worth mentioning is that even though the EER and min-tdcf is relatively high for this model as compared to the fusion model on the development set, it captures more generalized features than the fusion model, this can be confirmed by comparing the results of the development set and eval set for both the models.

\begin{figure}
    \centering
    \includegraphics[scale = 0.2]{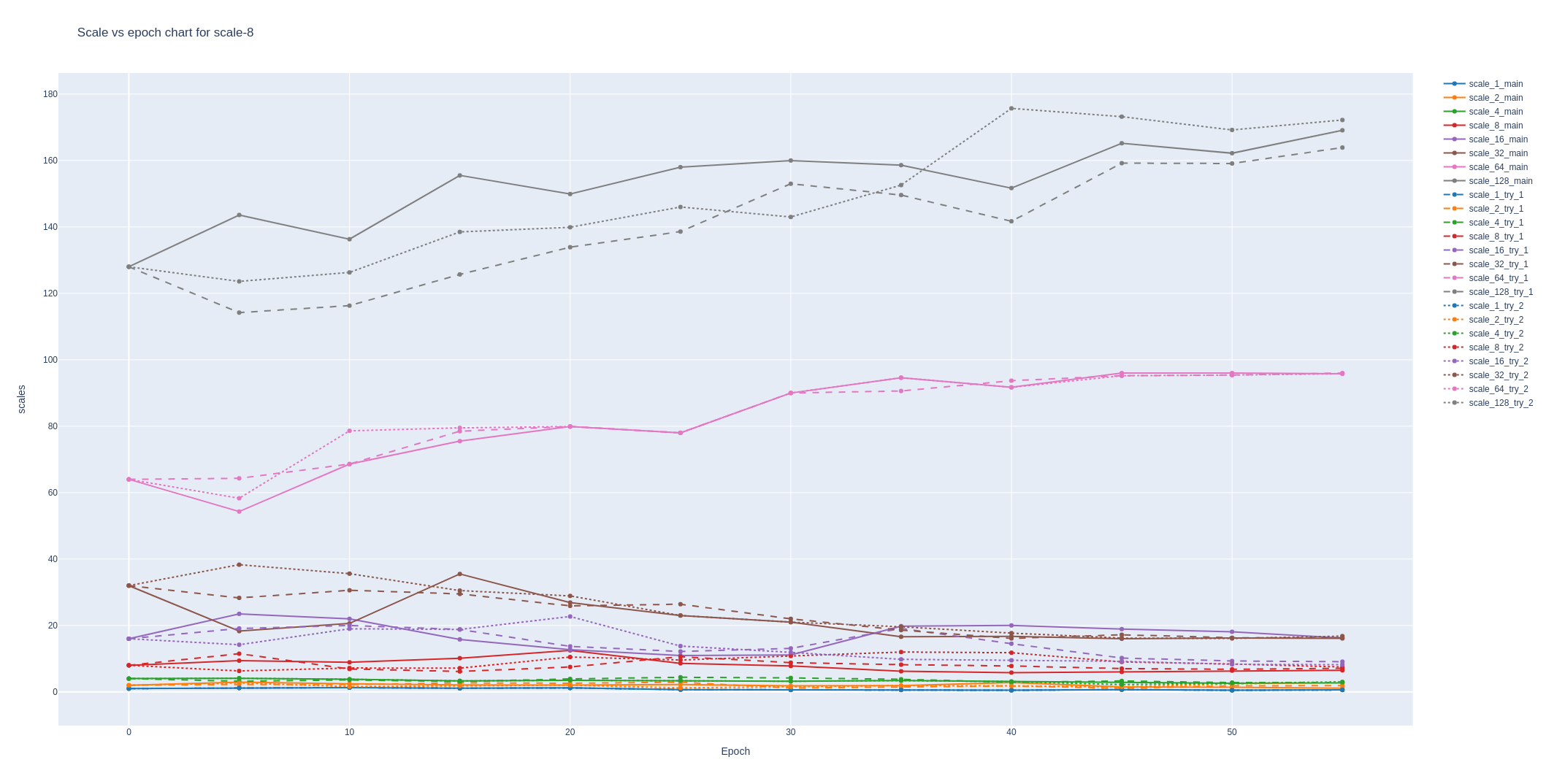}
    \caption{The visualizations of the scales versus epoch chart for the 8 scale wavelet deconvolution layer}
    \label{fig:scalevsepoch}
\end{figure}

The improvement in performance is due the hyper-parameter tuning capabilities of the WD layer that optimizes the scales to capture the relevant frequencies responsible for spoof detection. The Figure \ref{fig:scalevsepoch} is a scale vs epoch chart for the 8 scale wavelet deconvolution layer. It shows how the scale values change over the training process and at which scale values the model gives the best performance. And the model was trained 3 times to see if the final scale distribution was consistent each time. It can be observed that the scales initialized at 1,2,4,8,16,32 tend to converge towards lower scale values that is at higher frequency regions whereas the scales 64 and 128 converge towards higher constrained scale values. And this is congruent for all the three experiments.

% The visualizations of the 8 and 20 wavelet scales and the frequency response of the created filter-bank using the wavelet filters is given in the figure \ref{fig:Wavdeconv scale and frequency response}. It can be observed from the figure the scales are present mostly present in the upper and lower regions of the graph and from \ref{freqtoscale} it can be said that frequency is inversely proportional to scale this means that the relevant frequencies responsible for spoof detection are in higher frequency region and some in lower frequency region. 

% \begin{figure}[h]
%   \centering
%   \caption{Histogram and min-tdcf curve for wavelet deconvolution layer. Top row is for WavDeconv-8, second row is for WavDeconv-20}
%   \begin{minipage}[b]{0.3\textwidth}
%     \includegraphics[width=\textwidth]{Figures/deconv-hist-8.png}
%   \end{minipage}
%   \hfill
%   \begin{minipage}[b]{0.6\textwidth}
%     \includegraphics[width=\textwidth]{Figures/WD_min_tdcf.png}
%   \end{minipage}
%   \begin{minipage}[b]{0.3\textwidth}
%     \includegraphics[width=\textwidth]{Figures/deconv-hist-20.png}
%   \end{minipage}
%   \hfill
%   \begin{minipage}[b]{0.6\textwidth}
%     \includegraphics[width=\textwidth]{Figures/WD_min_tdcf-20.png}
%   \end{minipage}
% \label{fig:Wavdeconv histogram and mintdcf}
% \end{figure}

% - importance of feature selection 

% - difficulty in hyper-parameter tuning

% - automating the process of hyper-parameter tuning using back propagation proposed by khan and yener.

% - Performance of CNN as compared to the parametric filter approach

% - visualization of the learned scale parameter for the task of spoof detection. 

% - visualization of the frequency response  of the wavelet filters at learned scales.

% - difference in the learned scales parameters for task other than spoof detection to show that even though wavelets have limited success in the field of spoof detection this approach is able to given us an insight into the frequency regions that are important for the detection of spoofed speech.

\chapter{Conclusion} % Main chapter title

\label{Conclusion}

In this concluding chapter, I summarise my findings and arguments and reflect back on some of the strengths and weaknesses of the theoretical and methodological approaches I have taken in my progress towards building an End-to-End Parametric Learn-able Filter Approach. This thesis dissertation had three main experiments and each of these experiments helped build a case for the subsequent experiments that follow. And this finally directed us to build a wavelet based parametric learn-able filter approach.

The first experiment was carried out to test two things. The first was to compare the efficacy of MFCC and MWPC features to tackle spoofing attacks. And second was to test the reliability of the traditional countermeasure systems against modern state of the art spoofing attacks in the ASVspoof 2019 dataset. It was observed that MWPC due to its multi-resolutional analysis capabilities outperformed the MFCC features in tackling both traditional and modern spoofing attacks encountered in ASVspoof 2015 and 2019 dataset respectively. Furthermore, the traditional countermeasure performed pretty well for traditional HMM and Unit selection based spoofing attacks but their performance degraded severely for modern spoofing attacks. 

So in order to tackle the attacks that will be generated with upcoming voice conversion and TTS technologies we need a more robust and efficient countermeasure system that can surpass the performance of the traditional countermeasure systems. Based on the above observations and the superior efficacy of Wavelet transform our next experiments we developed and used, Sincnet which is our baseline and, CWTnet and WSnet, which are our wavelet based end to end Convolutional neural network based countermeasure systems that not only exploits the knowledge and expertise of signal processing through handcrafted features but also leverage the capabilities of the CNN to learn multiple levels of representation of data. The results show that While the performance of the standalone models were not encouraging it was seen that fusion with results from the Sincnet model provided significant improvements over the Sincnet baseline. It is apparent that these models are able to capture something significant that the Sincnet models seem to miss. 

The final scale distribution and number of scales used in CWTnet are far from optimal for the task at hand. This is where learn-able parametric filters come into the picture. The continuous transform layer in CWTnet was replaced with wavelet deconvolution layer which optimizes the scale parameter using back-propagation. It provides a fast and less computationally expensive method to do effective spectral decomposition as compared to other feature selection approaches. It was observed that wavelet deconvolution model performed much better then CWTnet and WSTnet, and its generalization capabilities surpassed all the models we previously proposed including the fusion results. Furthermore, the converged scale values suggested that low and high frequency regions were the most important for distinguishing between spoofed and genuine samples.

To ring down the curtains, traditional countermeasure systems are not reliable against modern neural network based spoofing attacks. The multi-resolutional analysis capabilities of wavelet transform seems to give it an edge over other handcrafted features when embedded in countermeasure systems to tackle spoofing attacks. On top of that wavelet based parametric learn-able filter layer showed an efficient way to perform feature selectivity and further improve the performance. The current deep learning based architectures used in our experiments are not able to truly exploit the potential that wavelet transform holds. And further investigation in more sophisticated architectures, loss functions and other multi-resolutional analysis based transforms can further improve the performance in spoof detection.

%----------------------------------------------------------------------------------------
%	THESIS CONTENT - APPENDICES
%----------------------------------------------------------------------------------------

\appendix % Cue to tell LaTeX that the following "chapters" are Appendices

% Include the appendices of the thesis as separate files from the Appendices folder
% Uncomment the lines as you write the Appendices

% \include{Appendices/AppendixA}
%\include{Appendices/AppendixB}
%\include{Appendices/AppendixC}

%----------------------------------------------------------------------------------------
%	BIBLIOGRAPHY
%----------------------------------------------------------------------------------------

\printbibliography[heading=bibintoc]

%----------------------------------------------------------------------------------------

\end{document}